\documentclass[preprint,square,numbers,sort&compress]{elsarticle}
\usepackage{setspace}
\usepackage{textcomp}
\usepackage[utf8]{inputenc}
\usepackage{multirow}
\usepackage{adjustbox}
\usepackage{xcolor}
\usepackage{float}
\usepackage[margin=1.0 in]{geometry}
\usepackage{wrapfig}
\usepackage{amsmath}
\usepackage[normalem]{ulem}
\usepackage{graphicx}
\usepackage{comment}
\usepackage{subcaption}
\usepackage{hyperref}

\begin{document}

\title{Picosecond Laser Ablation of Millimeter-Wave Subwavelength Structures on Alumina and Sapphire}

\author[1]{Qi Wen\corref{cor1}}
\ead{wenxx181@umn.edu}

\author[2]{Elena Fadeeva}

\author[1]{Shaul Hanany}

\author[2]{J\"{u}rgen Koch}

\author[3]{Tomotake Matsumura}

\author[4]{Ryota Takaku} 

\author[1]{Karl~Young}

\cortext[cor1]{Corresponding author}

\address[1]{School of Physics and Astronomy, University of Minnesota, Twin Cities, Minneapolis, USA}

\address[2]{Production and Systems Department, Laser Zentrum Hannover e.V., Hannover, Germany}
\address[3]{Kavli Institute for the Physics and Mathematics of the Universe (WPI),The University of Tokyo Institutes for Advanced Study, The University of Tokyo, Kashiwa, Chiba 277-8583, Japan}
\address[4]{Department of Physics, University of Tokyo, Tokyo, Japan}

\begin{abstract}

We use a 1030~nm laser with 7~ps pulse duration and average power up to 100~W to ablate pyramid-shape subwavelength structures (SWS) on alumina and sapphire. The SWS give an effective and cryogenically robust anti-reflection coating in the millimeter-wave band. We demonstrate average ablation rate of up to 34~mm$^3$/min and 20~mm$^3$/min for structure heights of 900~\textmu m and 750~\textmu m on alumina and sapphire, respectively. These rates are a factor of 34 and 9 higher than reported previously on similar structures. We propose a model that relates structure height to cumulative laser fluence. The model depends on  the absorption length $\delta$, which is assumed to depend on peak fluence, and on the threshold fluence $\phi_{th}$. Using a best-fit procedure we find an average $\delta = 630$~nm and 650~nm, and $\phi_{th} = 2.0^{+0.5}_{-0.5}$~J/cm$^2$ and $2.3^{+0.1}_{-0.1}$~J/cm$^2$ for alumina and sapphire, respectively, for peak fluence values between 30 and 70~J/cm$^{2}$.  With the best fit values, the model and data values for cumulative fluence agree to within 10\%.
Given inputs for $\delta$ and $\phi_{th}$ the model is used to predict average ablation rates as a function of SWS height and average laser power.

\end{abstract}


\maketitle

\section{Introduction}

For a wide range of millimeter and sub-millimeter (MSM) astrophysical instruments there is a need for broad-band, cryogenically robust, anti-reflection coatings (ARC)~\cite{adv_actpol,BA, polarbear2_instrument, SO, ali_china, s4_cdt,spt-3g-year2,alma_band1_design, litebird_spie2018}. Fabricating subwavelength structures (SWS) directly on the material of interest is an ARC technique that alleviates the need for several layers of materials and glues with various indices of refraction. The method is robust to cryogenic cycling because the layer of SWS is fabricated on the native element material, eliminating mis-matched coefficients of thermal expansion. SWS ARC have been used in a wide range of applications at various wavelengths; see Raut et al.~\cite{C1EE01297E} and references therein.

Alumina and sapphire ($\alpha$-Al$_2$O$_3$) have appealing properties as optical elements in the MSM bands. Both materials have low loss ($\tan \delta < 10^{-4}$) at cryogenic temperatures, thermal conductivities that are 1000 times larger than plastics at 77 K, and thermal contraction less than $0.1\%$~\cite{jacob2002microwave,Heidinger_1990,Molla_1993,pobell,ventura2014thermal,choy1981thermal}. With index of refraction near $n=3$, an alumina lens can be made thinner by at least a factor of 2 and maintain the same aberration correction power compared to a plastic lens that has $n=1.5$. 
A-cut sapphire has 10\% birefringence and is commonly used as a half-wave plate material~\cite{lamb96, hanany05}.
However, both materials have hardness 9 on the Mohs scale, making standard machining of SWS challenging~\cite{machining_difficult,Araujo2016}. They are also chemically inert and therefore chemical etching is not efficient~\cite{alumina_inert,sapphire_inert}.

In several previous publications we demonstrated that laser ablation can be used to fabricate pyramid-shape SWS ARC on alumina, sapphire, and silicon for MSM applications~\cite{matsumura2016_ARC, schutz2016,tomo_ISSTT, ryota_2019ieee,Ryota_2020,young_silicon}. 
We used pico- and femto-second lasers to fabricate structures with heights between 580 and 2100~\textmu m, and grid spacing, also called `pitch', in the range of 180 to 540~\textmu m. The achieved aspect ratios, defined as $a\equiv$height/pitch, were up to 5.3~\cite{tomo_ISSTT}. Measured transmittances were  
higher than 97\% over a band between 43 and 161~GHz on a sapphire sample~\cite{Ryota_2020}, and higher than 95\% over a band between 210 and 490~GHz on a silicon sample~\cite{young_silicon}.
The transmittances agreed with predictions based on the shape measurements of ablated structures, and calculations suggest that higher transmittance over a broader bandwidth is achievable upon optimization of the ablated structures~\cite{grann95}. 

The implementation of laser-ablated SWS ARC on alumina and sapphire for current and next generation instruments has been limited by slow fabrication rate. 
Optical elements including lenses, filters, and half-wave plates in several instruments have diameters near or larger than 500~mm~\cite{BA, SO, ali_china,s4_cdt}. Previously we reported volume removal rate of 1.0~mm$^{3}$/min on alumina and up to 2.2~mm$^{3}$/min on sapphire~\cite{matsumura2016_ARC,Ryota_2020}. With a volume removal rate of 2~mm$^{3}$/min, it would take more than 2~months of 24-hour ablation to fabricate a 1~mm-tall SWS ARC on both sides of a 500~mm-diameter optical element. Other researchers report rates between few hundredths 
to $\sim$10~mm$^{3}$/min on alumina and sapphire with a variety of fabrication shapes using short-pulse lasers~\cite{PERRIE2005213, CHEN2005214, Furmanski, Preusch,Engelhardt,ESMAIL2021106669}. Much higher rates, near 130~mm$^{3}$/min, have been reported with alumina for ablating shallow flat cavities using a high average power picosecond laser (up to 187 W) and a high-speed polygon scanner~\cite{alumina_fast_ablation}. A primary  goal of our research is to achieve SWS ARC ablation rates that are at least ten times faster than previously demonstrated. With rates near 20~mm$^3$/min, fabrication time of an ARC on a 500~mm diameter optical element would be reduced to within a week.

The dispersion in volume removal rates highlights the importance of modeling the ablation process and of establishing relations between laser parameters, fabricated shapes, and ablation rates.  When modeling the interaction between ultrashort pulses and dielectric materials, single- or multiple-rate-equation models are useful for describing the temporal evolution of free electrons in the material; see Balling and Schou~\cite{Balling_2013} and Rethfeld et al.~\cite{Rethfeld_2017}, and references therein. Hydrodynamic and molecular dynamic simulations provide insights into the mechanism of laser induced material changes (Ibid.). 
Limited computational resources make the implementation of these models prohibitive for simulating the end-to-end production of SWS ARC.  

A more suitable approach is to use a model appropriate for an industry-scale high-throughput ablation~\cite{Furmanski, Raciukaitis09,NEUENSCHWANDER20141047,Boerner_Hajri_2019}. Such a model necessarily includes simplifying assumptions, but has been shown to produce results that fit well with experimental data for both metals and dielectrics. 

Dispersion among reported ablation rates also arises from differences between experimental setups used in making such measurements. The ablation time counted in a pulse-by-pulse experiment includes mostly, if not exclusively, time in which the laser interacts with the surface; the ablation rate is indicative of the underlying physics. However, in experiments like the ones reported here, in which ``process time" includes the {\it entire} time to fabricate the sample, the reported rate could be affected by inefficiencies in the laser beam scan pattern across the surface. In such a case, the ablation rate is indicative of a combination of the underlying physics, {\it and} scan design inefficiencies. 

In this paper we report a set of laser ablation tests for fabricating pyramid-shape SWS with heights up to 1.1~mm and aspect ratio $a\sim 2.7$. We ablated alumina and sapphire with an IR picosecond laser and average power up to 100 W. We report on the measured material removal rate making a distinction between the `ablation rate' and the `process rate'. We extend an ablation model, first proposed by Furmanski et al.~\cite{Furmanski} and expanded upon by others~\cite{Raciukaitis09, NEUENSCHWANDER20141047}, and provide a relation between the height of the SWS and the cumulative fluence required to achieve that height.  The model depends on three parameters that we best-fit using the data: the threshold fluence $\phi_{th}$, and two parameters quantifying the absorption length $\delta$, which is assumed to depend on peak fluence. We fit for the three parameters and compare the model to the data. 

In Section~\ref{section:experiment} we give details about the laser, the scan parameters used to fabricate the SWS, optimization of the focus position relative to the surface, and the ablation trials. Results about the geometry of fabricated structures, process efficiency, and average ablation rate are provided in Section~\ref{sec:results}. In Section~\ref{sec:model} we describe the model for the ablation and compare it to results. Discussion and conclusions are given in Sections~\ref{sec:discussion} and~\ref{sec:conclusion}, respectively.

\section{Experimental setup}
\label{section:experiment}

\subsection{Laser Parameters and Sample Fabrication} 
\label{sec:samplefab}

\begin{figure}[h]
\centering
	\begin{subfigure}[t]{0.45\textwidth}
		\centering
		\includegraphics[width=1.0\textwidth]{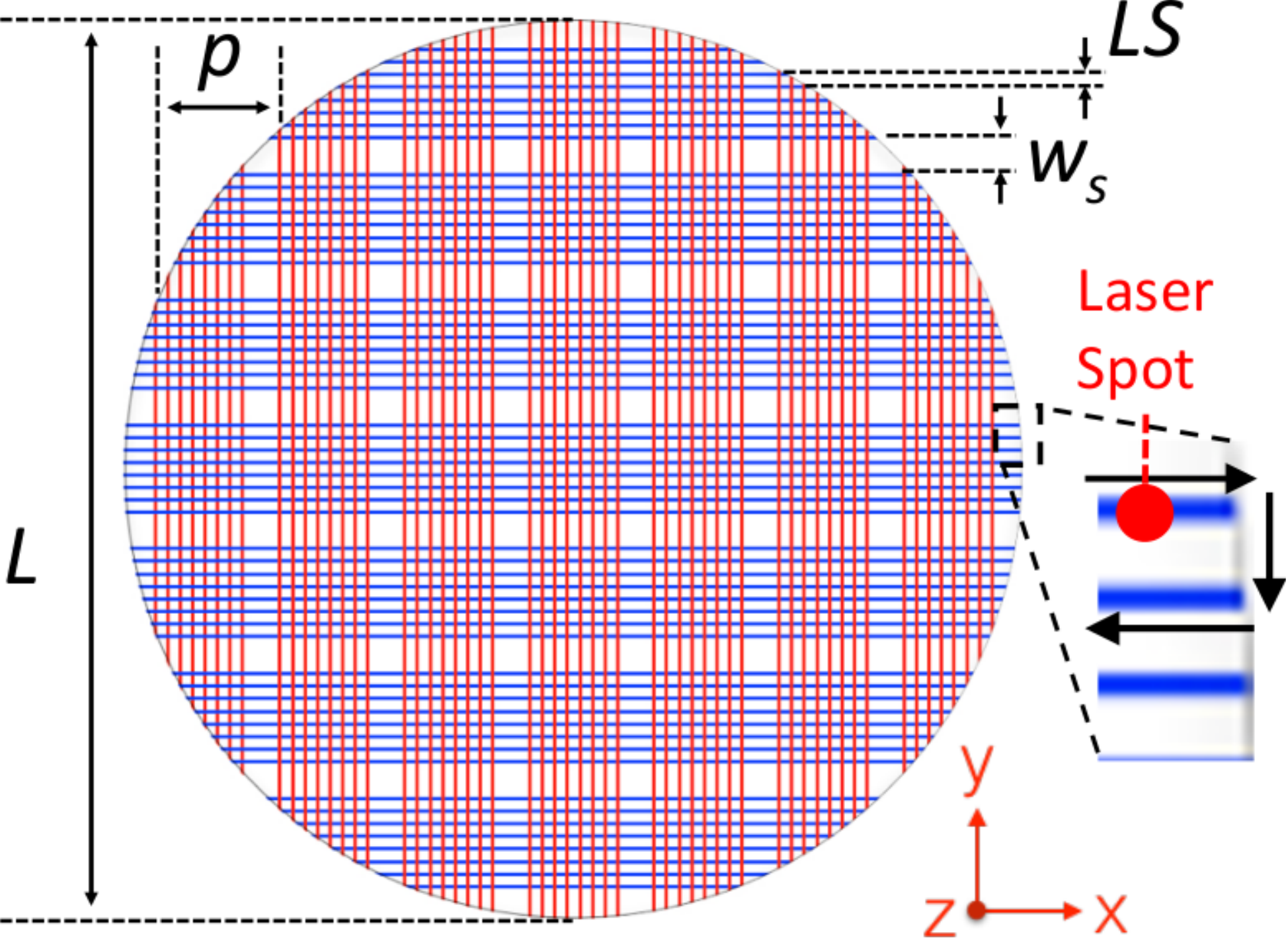}
		\caption{}		
	\end{subfigure}%
	\hspace{5mm}
		\begin{subfigure}[t]{0.45\textwidth}
		\centering
		\includegraphics[width=1.0\textwidth]{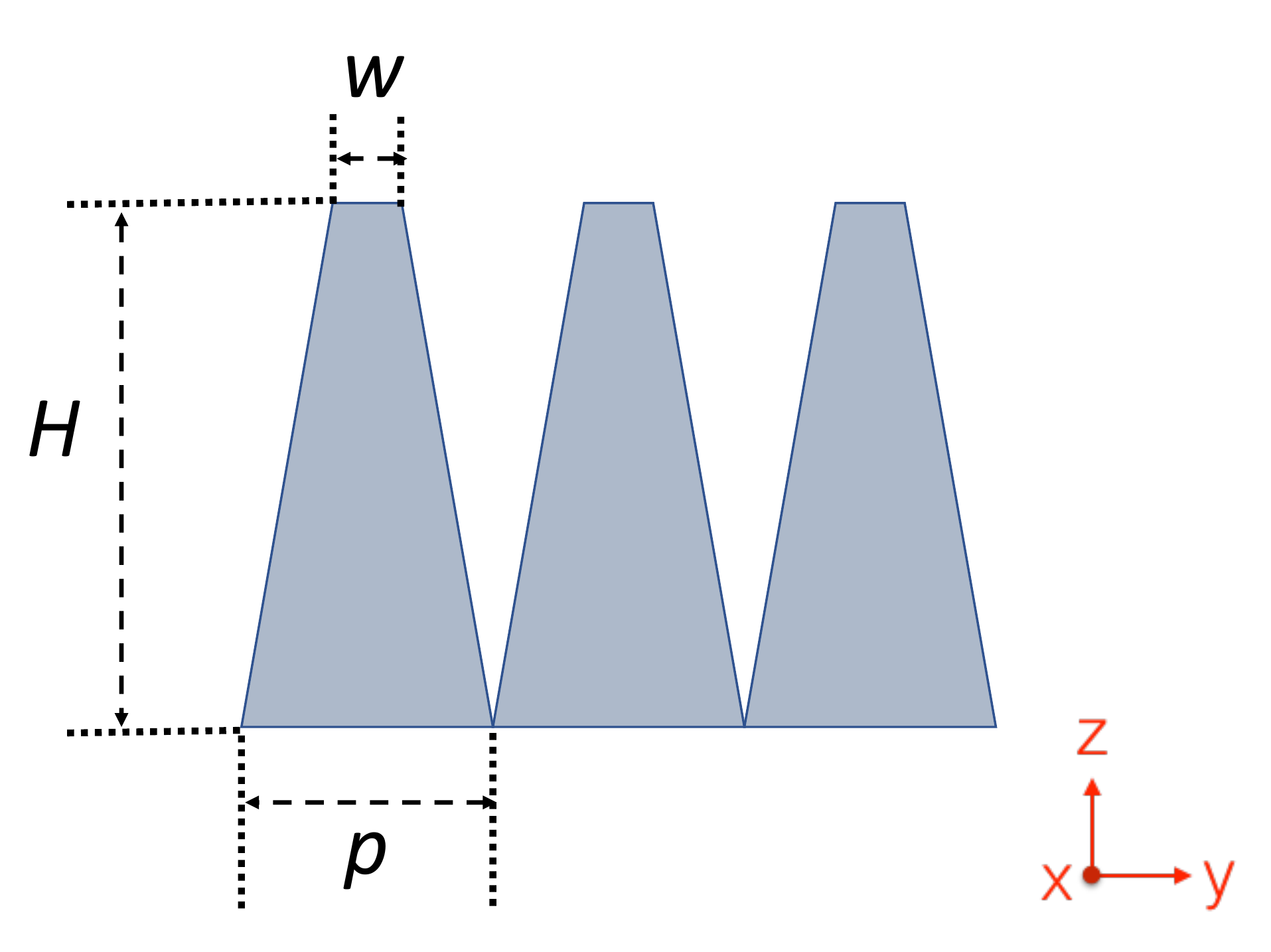}
		\caption{}		
	\end{subfigure}%
\caption{\textbf{(a)} The scan pattern used to fabricate the SWS. Each line represents a scan of the laser beam. The beam scanned all parallel lines along one axis (e.g. $y=$ red) then the other; this is one layer.  The laser transitions between lines by taking shortest path possible, thus adjacent line scans are opposite each other; see enlargement. The SWS are formed by repeating this scan for $N_{L}$ layers. Repeated ablation of lines spaced by $LS$ creates grooves with pitch $p$ over a circular or square area with side length $L$. The scan parameters were varied and are given in Table~\ref{tab:scan}. \textbf{(b)} Side-view sketch of the design of the SWS. The structure head size $w$ is designed to be smaller than $w_s$; compare Table~\ref{tab:scan} and Section~\ref{sec:fabricated_structures}.
\label{fig:scan} }
\end{figure}

We fabricated SWS on one side of flat discs of alumina and sapphire using a Trumpf TruMicro 5070 picosecond laser. A jet of compressed air removed ablation debris during processing, and the samples were cleaned in an ultrasonic bath. We used a 100~mm focal-length lens to focus the laser beam, which had a diameter of 5.2~mm at the lens. The other laser parameters used are given in Table~\ref{tab:lasers}. The SWS were formed by repeating a specific scan pattern of the laser beam $N_{L}$ times across the sample. We refer to each repeat as a `layer' and thus a full fabrication consisted of $N_{L}$ layers. A scan pattern and a sketch of the side view of SWS are shown in Figure~\ref{fig:scan}, and the parameters of scan patterns are given in Table~\ref{tab:scan}.
\begin{table}[h]
    \centering
    \normalsize
    \begin{tabular}{|c|c|}
    \hline
Pulse duration [ps] & 7  \cr     \hline
Wavelength [nm] & 1030   \cr     \hline
Max average power [W]& 100   \cr    \hline
Repetition rate [kHz] & 400, 600   \cr    \hline
Focal spot diameter [\textmu m] & 28   \cr    \hline
Rayleigh length [\textmu m] & 538  \cr    \hline
    \end{tabular}
    \caption{Laser parameters used in the ablation experiments. The focal spot diameter is where the intensity drops to $1/e^2$ of the peak intensity. The set of specific ablation trials is listed in  Table~\ref{tab:trials}.
    \label{tab:lasers} }
\end{table}

A layer was made up of groups of closely spaced lines that were laid on a grid with pitch $p$. Within a group, lines were separated by distance $LS$, and there were $n_{lines}$ per group. Ablation of the closely grouped $n_{lines}$ created a groove. The combination of all grooves made the pyramid structures. The laser scanned all $y$-direction lines then all $x$-direction lines to give one layer. The number of layers for each ablation trial is given in Table~\ref{tab:trials}. At the end of the scan of a line the laser transitions to the adjacent line, or to the adjacent groove, by taking the shortest path possible such that successive line are scanned in opposite directions; see Figure~\ref{fig:scan}.

\begin{table}[!h]
    \centering
    \normalsize
    \begin{tabular}{|c|c|c|c|c|c|c|c|}
    \hline
    Scan Pattern & $p$ [\textmu m] & $LS$ [\textmu m] & $w_s$ [\textmu m] & $n_{lines}$  & $L$ [mm] \\ \hline
    \#1  & 400 & 40 & 120 & 8 & 2.85 \\ \hline
    \#2  & 330 & 30 & 150 & 7 & 2.46 \\ \hline
    \end{tabular}
    \caption{ 
   Scan patterns used to fabricate the SWS. They all followed the pattern shown in Figure~\ref{fig:scan}.
    \label{tab:scan} }
\end{table}

The laser focus position in $z$ was set at the beginning of fabrication and kept constant throughout. It was set at -0.75~mm, with negative values signifying a position below the original surface of the material. This focus position, which is sometimes called defocus or defocusing distance~\cite{defocus1,defocus2,defocus3},
had been optimized through experimentation as described next.

\subsection{Focus Position Optimization}
\label{sec:focus}

\begin{figure}[H]
\centering
	\begin{subfigure}[t]{0.5\textwidth}
		\centering
		\includegraphics[width=1.0\textwidth]{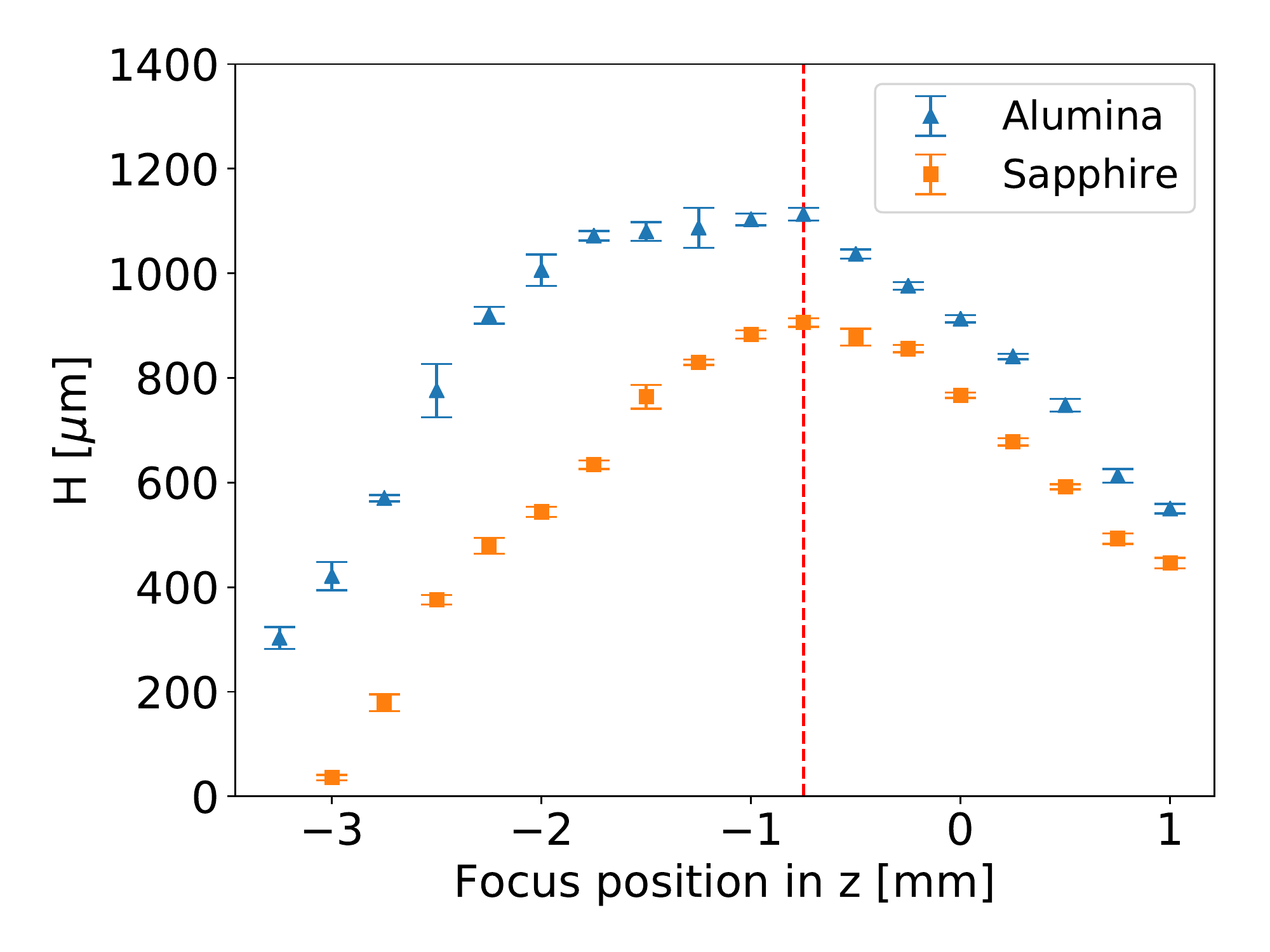}
		\caption{}		
	\end{subfigure}%
		\begin{subfigure}[t]{0.5\textwidth}
		\centering
		\includegraphics[width=1.0\textwidth]{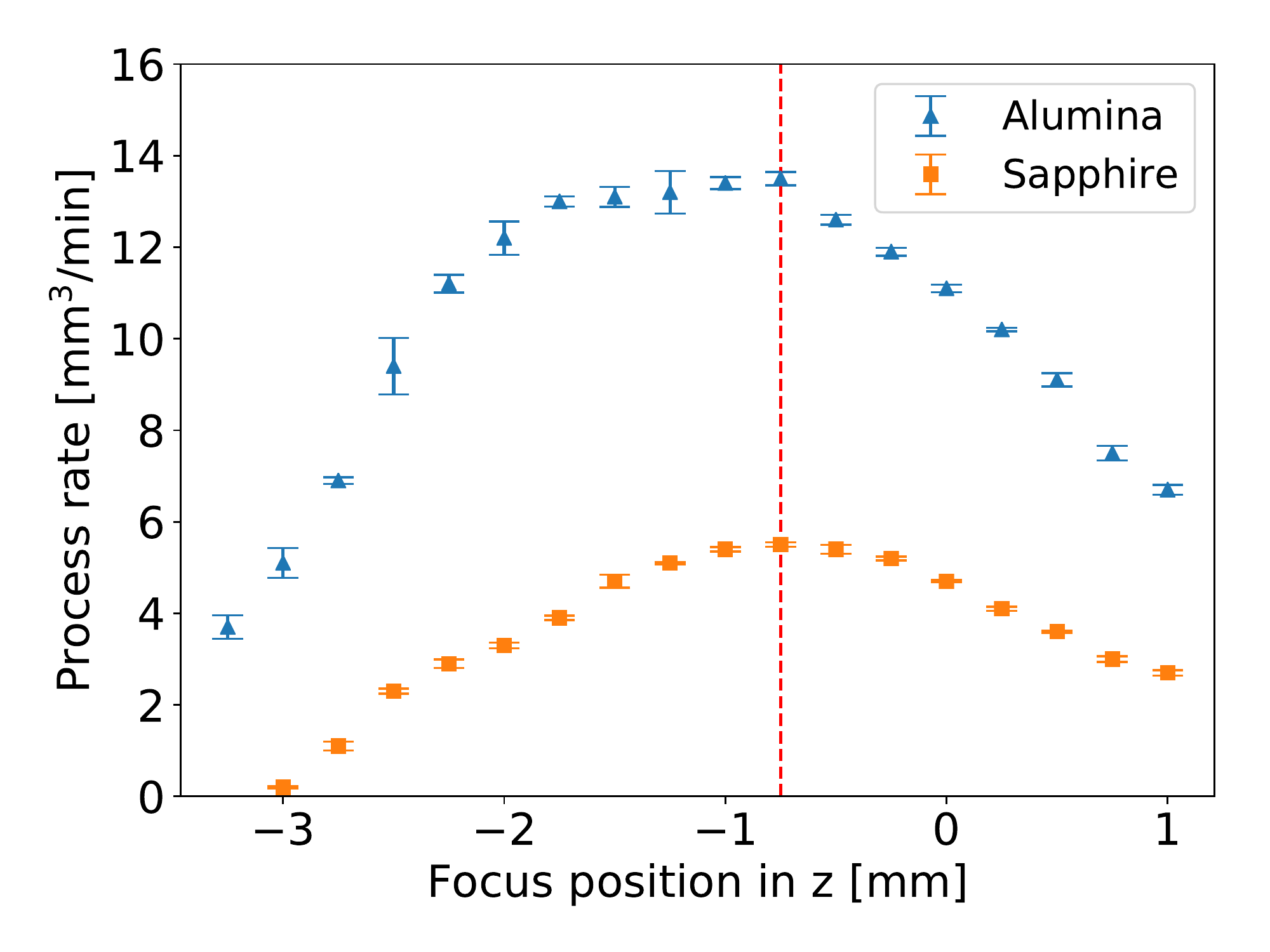}
		\caption{}		
	\end{subfigure}%
\caption{Final SWS height \textbf{(a)} and process rate \textbf{(b)}, as a function of laser focus position. The error bars are the standard deviations from measurements of several pyramids at each focus position. Each measurement is limited by the 5~\textmu m accuracy of the microscope. Negative focus positions indicate focus below the surface. The terminology `process rate' is explained in Section~\ref{sec:removal_rate}. We chose a focus position of -0.75~mm (vertical dash line) for the trials listed in Table~\ref{tab:trials}.
\label{fig:focus} }
\end{figure}

With all other parameters fixed, as detailed in Table~\ref{tab:focus_optimization}, we repeated ablation of both alumina and sapphire while varying the focus position $z$ between +1 and -3~mm. For both materials, we found that positioning the beam focus 0.75 to 1.00~mm below the surface gave taller structures and faster volume removal rate compared to other $z$ positions; see Figure~\ref{fig:focus}. The results indicate a relatively broad optimum for alumina, with values between $z=-0.50$~mm and $z=-2.00$~mm giving results that are within 10\% of the maximum. For sapphire, the same range spans values between $z=-0.25$~mm and $z=-1.25$~mm. These results are reasonable given that with negative focus more of the beam energy was confined below the surface. The remainder of ablation trials reported  below were conducted with a {\it fixed} focus position at $z=-0.75$~mm.

\begin{table}[!h]
    \centering
    \begin{tabular}{|c|c|}
    \hline
Power  & 50 W  \cr     \hline
Repetition rate  & 400 kHz  \cr     \hline
Focus position z & Between -3~mm and +1~mm  \cr     \hline
Scan for alumina & Scan pattern \# 1$^{a}$, scan speed $v_s = 0.50$ m/s, number of layers $N_L=30$ \cr    \hline
Scan for sapphire & Scan pattern \# 2, scan speed $v_s = 1.00$ m/s, number of layers $N_L=80$ \cr    \hline
\multicolumn{2}{l}{ $^{a}$ {\small With these modifications: $p =$ 370~\textmu m, $w_s =$ 90~\textmu m} \normalsize} 
    \end{tabular}
    \caption{Parameters for focus position optimization. Parameters not listed here have been fixed at the values given in Tables~\ref{tab:lasers} and~\ref{tab:scan}.  
    \label{tab:focus_optimization} }
\end{table}

\subsection{Summary of Tests}
\label{sec:summaryoftests} 

We conducted ablation tests in which we varied the number of layers, the laser scan speed, the pulse repetition rate, and pulse energies as given in Table~\ref{tab:trials}. We focused on varying pulse energy. For most pulse energies, as allowed by constraints of total power, we tested four configurations of $N_{L}$, repetition rate, and scan speed, adjusting the last two such that 
the total energy delivered per layer only depended on pulse energy.

We recorded the total process time for each trial and  post-fabrication imaged the samples with an optical microscope to measure the geometry of the SWS. For each sample four pyramids were measured. 

\begin{table}[!h]
\small
    \centering
    
    \begin{tabular}{|c|c|c|c|c|c|c|c|}
    \hline
     &  Trial    & Scan & Number of  & Scan speed $v_{s}$  & Repetition rate & Pulse energy & Average power $P$    \\
          &            & Pattern    & layers $N_{L}$ &  (m/s) &      (kHz) &    (\textmu J) &  (W)   \\
    \hline
    
     {\multirow{4}{*}{\rotatebox[origin=c]{90}{Alumina}}} & 1   & \#1 & 15 & 0.50  & 400 & 75, 100 ..., 225, 250$^{a}$ & 30, 40, ..., 90, 100  \\ \cline{2-8}
      & 2 & \#1 & 30 &  0.50  & 400  & 75, 100, ..., 225, 250$^{a}$ & 30, 40, ..., 90, 100   \\ \cline{2-8}
     & 3  & \#1 & 15 &  0.75 & 600  & 75, 100, 125, 150, 166 & 45, 60, 75, 90, 100  \\ \cline{2-8}
      & 4 & \#1 & 30 &  0.75 & 600  & 75, 100, 125, 150, 166 & 45, 60, 75, 90, 100  \\ 
    \hline
    \hline
     {\multirow{4}{*}{\rotatebox[origin=c]{90}{Sapphire}}} & 5 & \#2 & 40 & 1.00 & 400 & 75, 100 ..., 225, 250$^{a}$ & 30, 40, ..., 90, 100  \\ \cline{2-8}
      & 6  & \#2 & 80 &  1.00 & 400  & 75, 100 ..., 225, 250$^{a}$ & 30, 40, ..., 90, 100 \\ \cline{2-8}
     & 7  & \#2 & 40 &  1.50 & 600  & 75, 100, 125, 150, 166 & 45, 60, 75, 90, 100 \\ \cline{2-8}
      & 8  & \#2 & 80 &  1.50 & 600  & 75, 100, 125, 150, 166 & 45, 60, 75, 90, 100 \\  \cline{2-8}

    \hline
    \multicolumn{8}{l}{ $^{a}$ {\small Pulse energy ranged between 75 and 250~\textmu J with 25~\textmu J increment. \normalsize }}
    \end{tabular}
    \caption{
    Ablation trials. For each trial we fabricated a number of samples each with the pulse energy listed. Trial 1, for example, produced eight samples. 
    \label{tab:trials} }
\end{table}

\section{Results}
\label{sec:results}

\subsection{Characteristics of Fabricated Structures and Ablation Process}
\label{sec:fabricated_structures}

\begin{figure}[h]
\centering
\includegraphics[scale=0.4]{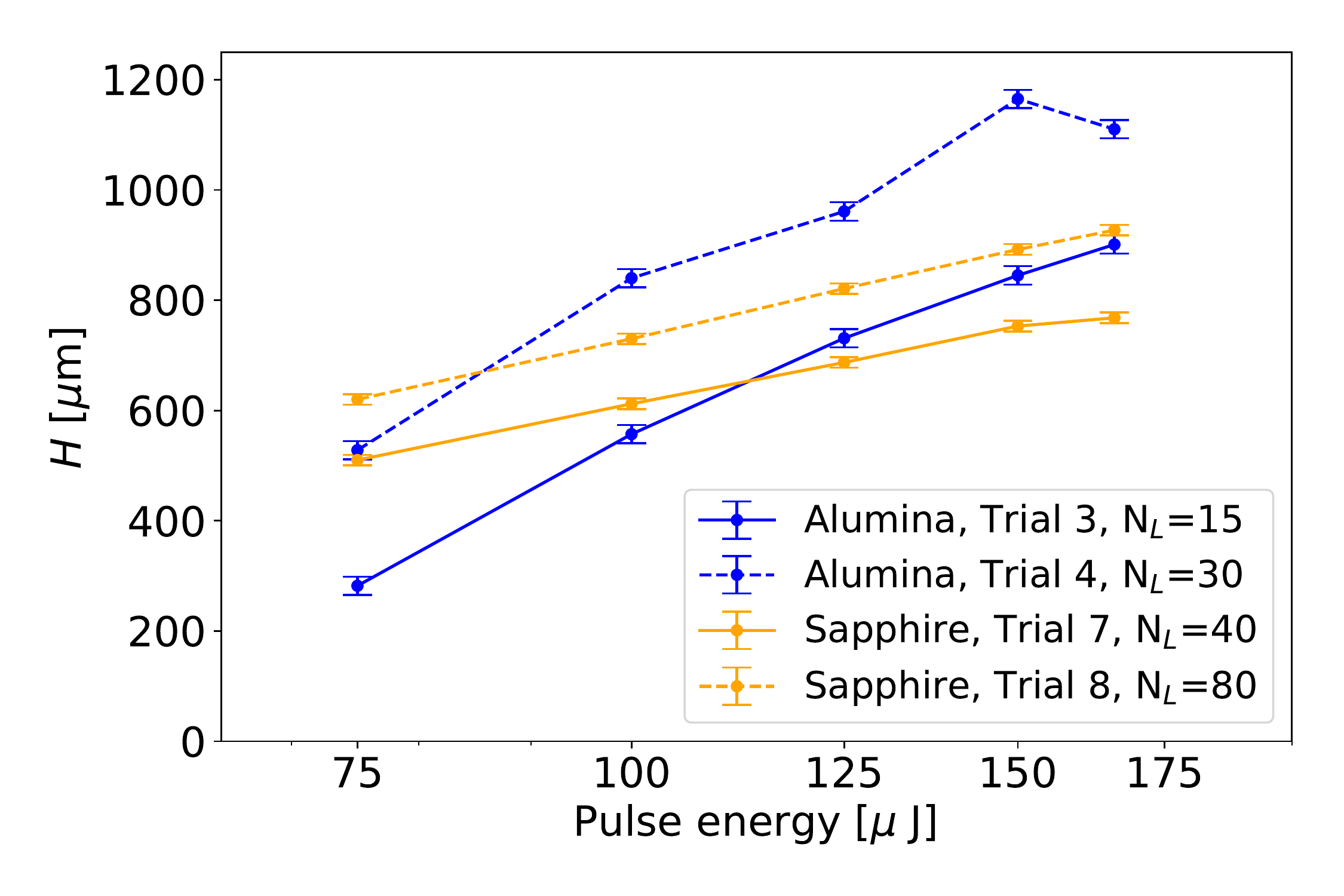}
\caption{Examples of SWS height as a function of pulse energy and the number of layers $N_{L}$. Higher pulse energy and higher $N_{L}$ both lead to taller structures. Lines are for visualization purpose; they connect points belonging to the same dataset
\label{fig:height_raw_data} }
\end{figure}

\begin{figure}[h]
\centering

	\begin{subfigure}[t]{0.3\textwidth}
		\centering
		\includegraphics[width=1.0\textwidth]{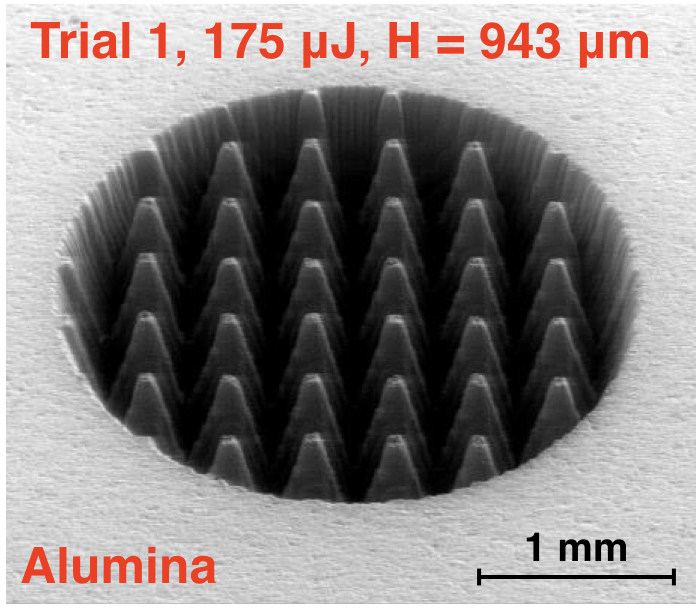}
		\caption{}		
	\end{subfigure}%
		\begin{subfigure}[t]{0.3\textwidth}
		\centering
		\includegraphics[width=1.0\textwidth]{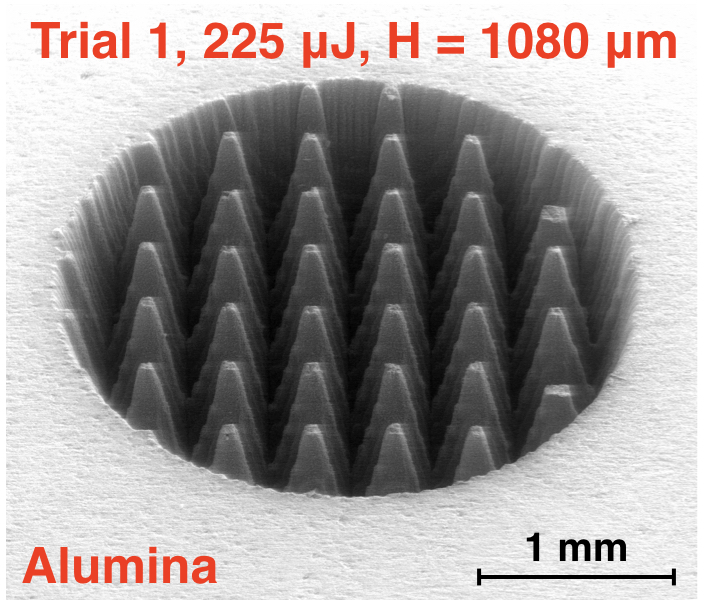}
		\caption{}		
	\end{subfigure}%
		\begin{subfigure}[t]{0.3\textwidth}
		\centering
		\includegraphics[width=1.0\textwidth]{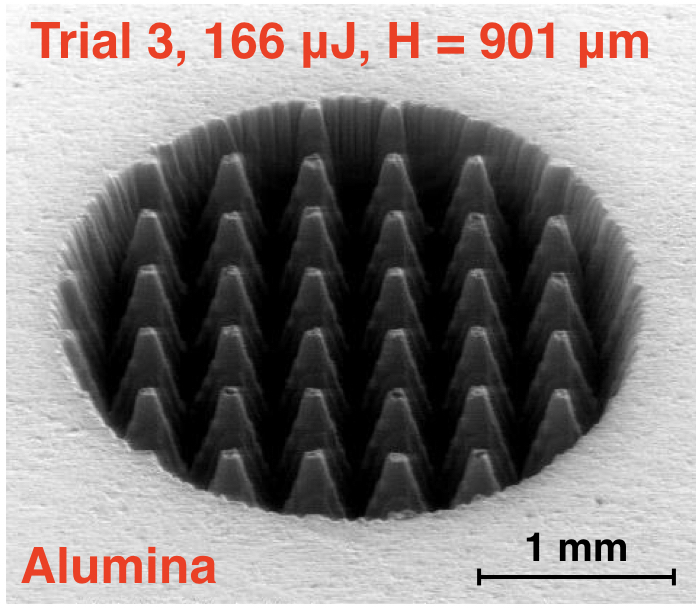}
		\caption{}		
	\end{subfigure}%
	
		\begin{subfigure}[t]{0.3\textwidth}
		\centering
		\includegraphics[width=1.0\textwidth]{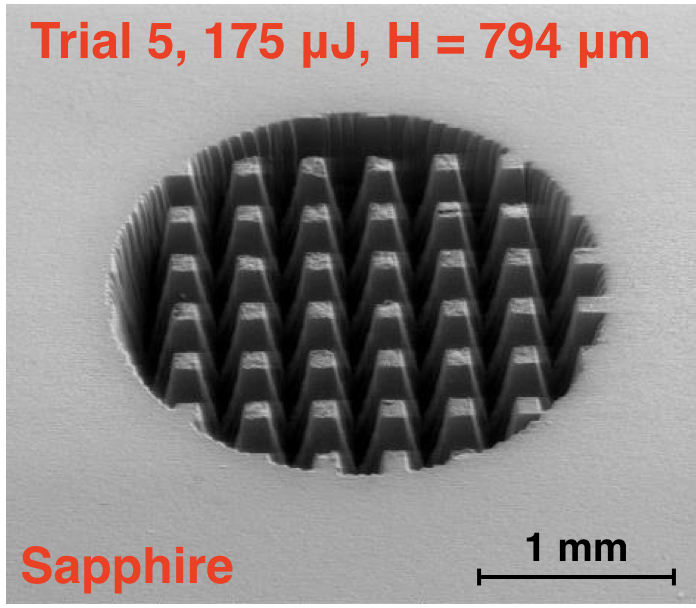}
		\caption{}		
	\end{subfigure}%
		\begin{subfigure}[t]{0.3\textwidth}
		\centering
		\includegraphics[width=1.0\textwidth]{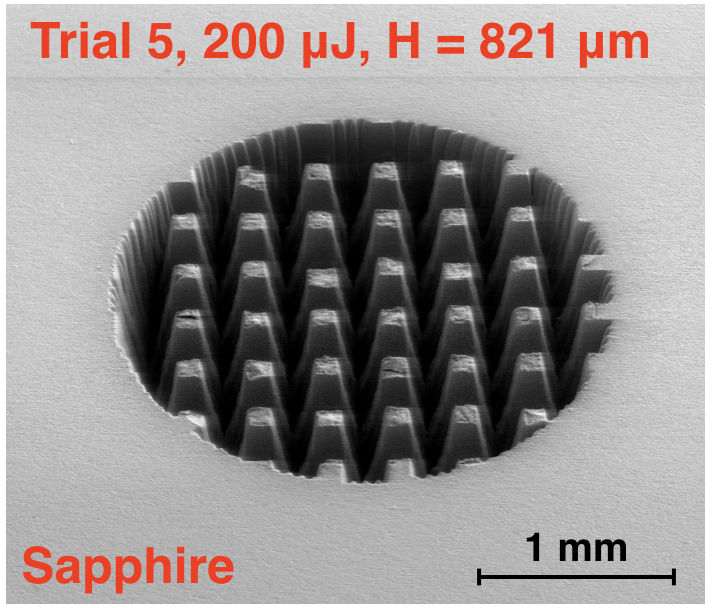}
		\caption{}		
	\end{subfigure}%
		\begin{subfigure}[t]{0.3\textwidth}
		\centering
		\includegraphics[width=1.0\textwidth]{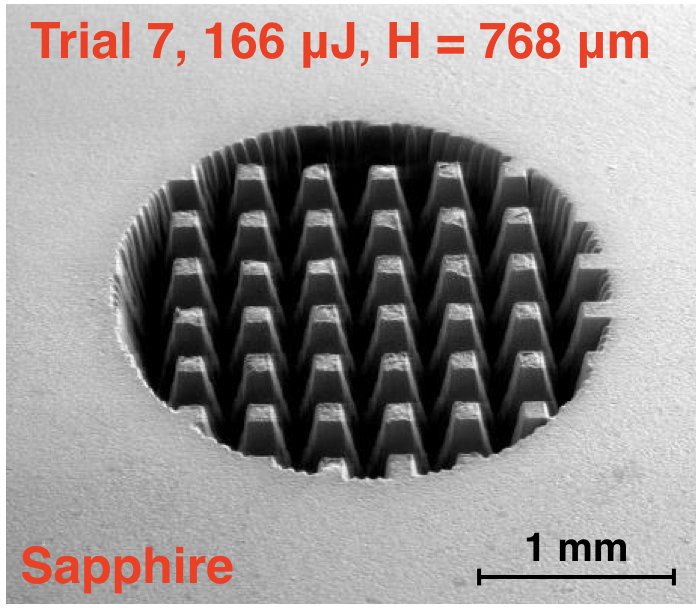}
		\caption{}		
	\end{subfigure}%
\caption{Photographs of a subset of the fabricated SWS samples. Alumina samples are shown in the top panels, and sapphire samples are shown in the lower panels.} 
\label{fig:elena_photos} 
\end{figure}

The scan pattern produces SWS, which in the majority of cases are intact pyramid-like shapes over the entire fabrication area; see Figure~\ref{fig:elena_photos}. The measured pitch agrees with the design values given in Table~\ref{tab:scan}. We find that the final height of the structures is a function of the pulse energy and the number of scan layers; see Figure~\ref{fig:height_raw_data}. We assign a height uncertainty of 17~\textmu m and 10~\textmu m for alumina and sapphire, respectively, based on the average values of the measured standard deviations shown in Figure~\ref{fig:focus}. In a small minority of cases, specifically for some of the highest energies and tallest structures, we find varying degree of damage to the pyramids, including breakage or cracking of tips. The results we report conservatively exclude all trials for which alumina (sapphire) SWS height is larger than 1100 (850)~\textmu m and two sapphire samples that would have been excluded by this criterion but had significant number of broken tips and thus gave an average height below 850~\textmu m. For the structures we report here, no pyramids are missing and the vast majority of pyramids are completely intact; Figure~\ref{fig:elena_photos} is representative. With these heights and the measured pitch the maximum aspect ratios are $a=2.75$ and 2.6 for alumina and sapphire, respectively.  The measured head size $w$ is correlated with  structure height such that $w$ is somewhat smaller for taller structures. Even with this correlation the standard deviation for $w$ is only 10\%. The average and standard deviation for all pyramids and all samples are $w = 70\pm7$~\textmu m on alumina and $82\pm 8$~\textmu m on sapphire. We use these averages in the ablation model discussed below.

For each trial we calculate the cumulative fluence $F_{cum} \equiv E/A_{a} = P T_{a}/A_{a}$, which is the ratio of the total energy delivered, $E$, to the ablated area, $A_{a}$. The energy $E$ is the product of average laser power and ablation time $T_a$; see more details about $T_a$ in Section~\ref{sec:efficiency}. In calculating $A_{a}$ we include only the regions where the laser beam interacts with the material, not the entire sample area $A_{sample}$ on which SWS have been fabricated; see Figure~\ref{fig:scan}. Ignoring edge effects
\begin{equation}
    A_{a} = A_{sample} (1-w^2/p^2) 
        = \Omega L^2 (1-w^2/p^2)
\label{eq:ablated_area}
\end{equation}
where the quantities $L, w, \, p$ have been defined in Figure~\ref{fig:scan}, and $\Omega$ is a form factor with values $\Omega=\pi/4$ for the circular samples we present here, or $\Omega=1$ for a square sample. For several of the samples we compared the actual measured area $A_{a}$ to the values obtained using Equation~\ref{eq:ablated_area} and found that they agree within 3\%, and we therefore use the analytical estimate for subsequent calculations.

\begin{figure}[h]
\centering
\begin{subfigure}[t]{0.5\textwidth}
		\centering
		\includegraphics[width=1.0\textwidth]{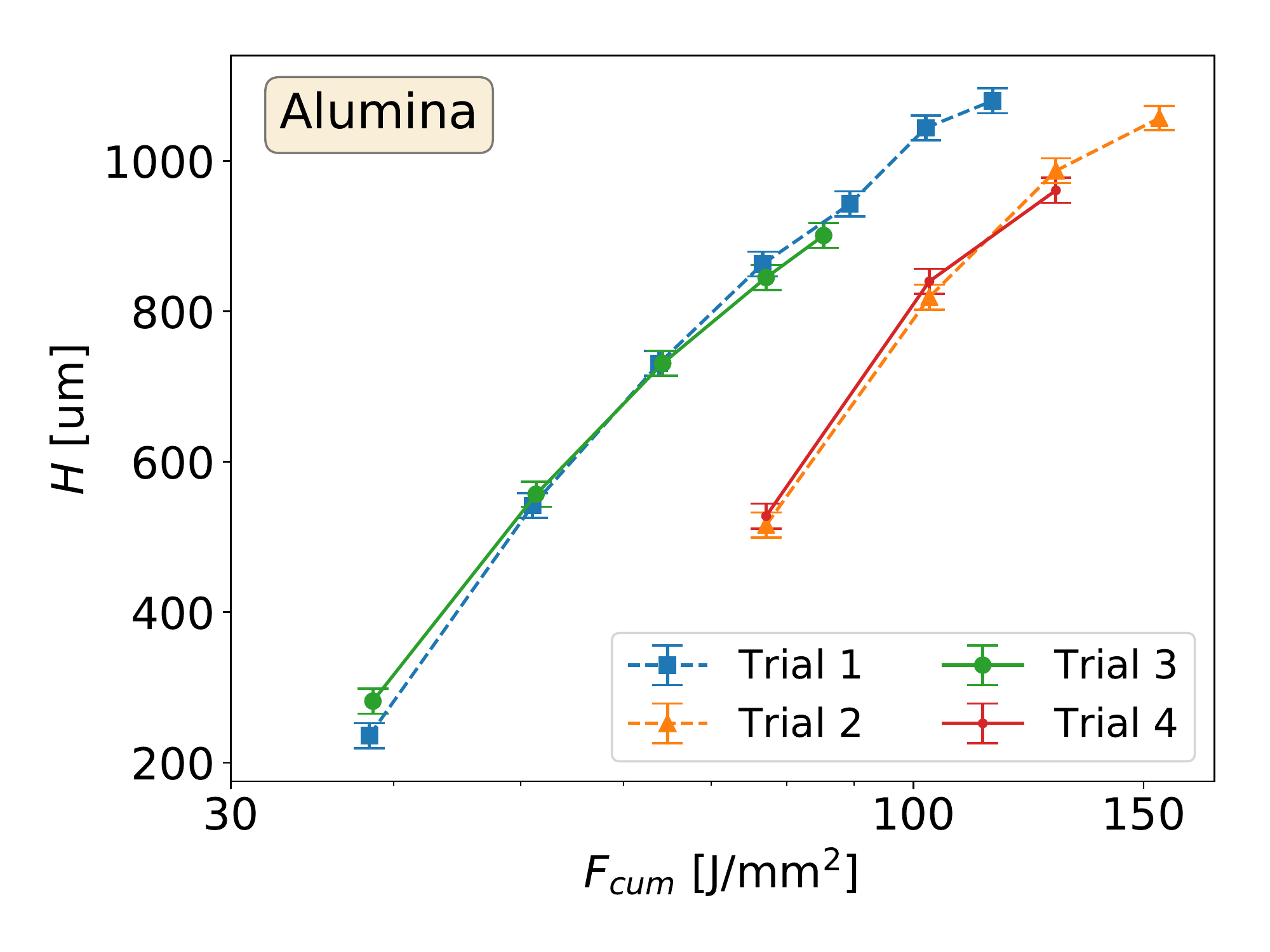}
		\caption{}		
	\end{subfigure}%
\begin{subfigure}[t]{0.5\textwidth}
		\centering
		\includegraphics[width=1.0\textwidth]{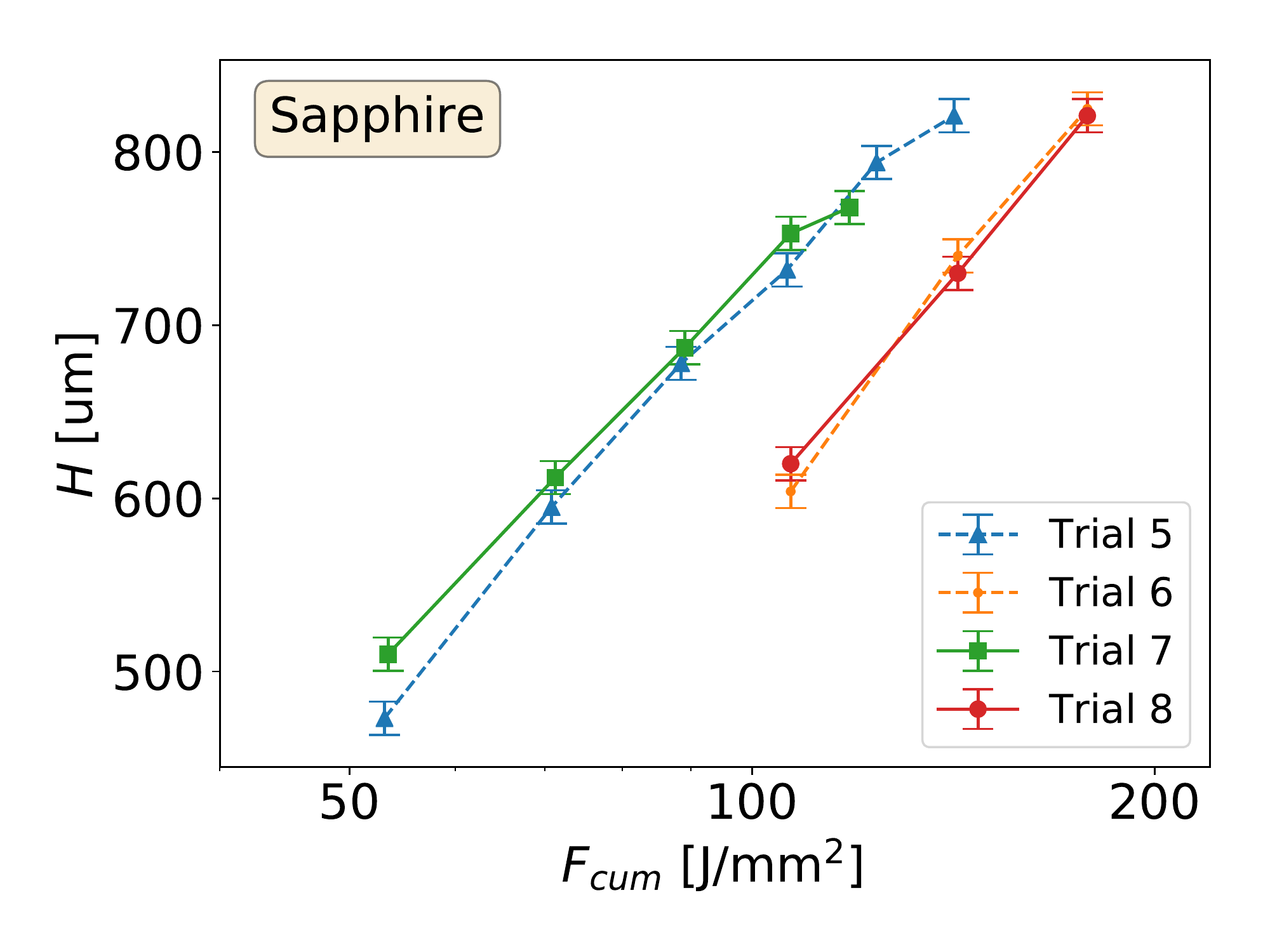}
		\caption{}		
	\end{subfigure}%
\caption{SWS height $H$ as a function of cumulative fluence $F_{cum}$ for the \textbf{(a)} alumina and \textbf{(b)} sapphire samples. 
\label{fig:elena_height_energy} }
\end{figure}

Within a given trial the height of the SWS is by-and-large a monotonically increasing function of $F_{cum}$; see Figure~\ref{fig:elena_height_energy}. The trial pairs (1,3), (2,4), (5,7), and (6,8), where the curves overlap, share the same $N_{l}$; the repetition rate and scan speed were adjusted to maintain the same $F_{cum}$.  The similarity between the curves suggest that a single underlying model may account for all the data; this is the topic of Section~\ref{sec:model}.

\subsection{Ablation Time and Process Efficiency}
\label{sec:efficiency}

The `process time' $T_{p}$ is the time it takes to complete the ablation in a given trial, or more generally in a given fabrication process. The process time includes the `ablation time' $T_{a}$ during which the laser ablates the material, and dead times during which the laser is off. Dead time includes periods $T_{trans}$ when the beam transitions between lines and other delays $T_{delay}$:
\begin{equation}
T_p = T_a + T_{trans} + T_{delay}.
\label{eq:tp}
\end{equation}
For a fixed SWS geometry, reducing $T_{p}$ requires reducing $T_{a}$ by optimizing material ablation parameters, and reducing non-ablation times $T_{trans}$ and $T_{delay}$ by improving process efficiency defined as $\epsilon \equiv T_{a}/T_{p}$. It is useful to make a distinction between $T_{a}$ and $T_{p}$ because improving $T_{a}$ requires understanding and optimizing the physics of the ablation process. When increasing efficiency one concentrates on scan strategy and on removing parasitic non-ablation times.  

In~\ref{sec:process_efficiency_derivation} we show that for the scan pattern described in Section~\ref{section:experiment} the ablation time is
\begin{equation}
T_a = 2 \Omega N_{L}  (L/p)  n_{lines}  (L/v_{s}),
\label{eq:ta}
\end{equation}
and that a time $\tau$ can be defined such 
\begin{equation}
\epsilon = \frac{T_{a}}{T_{p}} = \frac{T_{a}}{T_a + T_{trans} + T_{delay}} = 
  \frac{L/v_{s}}{L/v_{s}+\tau};  
\label{eq:efficiency2}
\end{equation}
The duration $\tau$ is an average parasitic (=non ablation) time per line that depends on scan parameters and hardware delay times but not on the sample size. 
For large samples, for which $L/v_{s}\gg\tau$, $\epsilon\approx1$. Tests conducted on small samples, especially those with fast scan speed, have lower process efficiencies due to a relatively larger contribution from parasitic processes. In Table~\ref{tab:efficiency}, we give the process and ablation times, the calculated process efficiency, and the inferred duration $\tau$ for each of the trials. Equation~\ref{eq:efficiency2} is not valid for scan strategies in which line scans are in the same direction. In such a scan strategy $\tau$ does depend on sample size, reducing $\epsilon$.

\begin{table}[!h]
    \centering
    \begin{tabular}{|c|c|c|c|c|c|}
    \hline
     & Trial & $T_p$ [s] & $T_a$ [s] & $\epsilon$ [\%] & $\tau$ [ms]  \\
    \hline
     {\multirow{4}{*}{\rotatebox[origin=c]{90}{Alumina}}} & 1 & $9.9\pm~0.1$ & $7.7\pm~0.2$ & $77\pm~2$   & $1.7\pm~0.2$\\ \cline{2-6}
     & 2 & $19.6\pm~0.1$ & $15.3\pm~0.5$ & $78\pm~2$ & $1.6\pm~0.2$ \\ \cline{2-6}
     & 3 & $7.1\pm~0.1$ & $5.1\pm~0.2$ & $72\pm~2$ & $1.5\pm~0.2$\\ \cline{2-6}
     & 4 & $14.4\pm~0.6$ & $10.2\pm~0.3$ & $71\pm~4$ & $1.6\pm~0.3$\\ \cline{2-6}
    \hline
    \hline
     {\multirow{4}{*}{\rotatebox[origin=c]{90}{Sapphire}}} & 5 & $13.7\pm~0.2$ & $8.1\pm~0.2$ & $59\pm~2$  & $1.7\pm~0.1$\\ \cline{2-6}
     & 6 & $27.3\pm~0.2$ & $16.1\pm~0.5$ & $59\pm~2$  & $1.7\pm~0.1$\\ \cline{2-6}
     & 7 & $10.5\pm~0.1$ & $5.4\pm~0.2$ & $51\pm~2$ & $1.6\pm~0.1$\\ \cline{2-6}
     & 8 & $21.1\pm~0.3$ & $10.8\pm~0.3$ & $51\pm~2$ & $1.6\pm~0.1$\\ \cline{2-6}
    \hline
    \hline
    \end{tabular}
    \caption{Summary of the experimental process and ablation times, process efficiencies, and derived parasitic times. We show the average of recorded $T_p$ with the standard deviation, since the same scanning was repeated over different pulse energies. The errors for $T_a$, $\epsilon$ and $\tau$ are the standard deviations after uncertainty propagation. 
    \label{tab:efficiency} }
\end{table}

\subsection{Average Ablation Rate}
\label{sec:removal_rate}

During ablation, material is removed at an average ablation rate $\overline{V_{a}}=\Delta V / T_a$, where $\Delta V$ is the volume of material removed during ablation time $T_{a}$. 
For short time intervals, the {\it instantaneous} removal rate $v_{rr}$ and  $\overline{V_{a}}$ are equal, but they have different values when considering the entire ablation process because $v_{rr}$ varies as the height of the ablated structures increases. We define the `process rate' as 
$\overline{V_{p}}=\Delta V / T_{p}$. 
For sufficiently high process efficiency it is possible to have $\overline{V_{p}} \simeq \overline{V_{a}}$.

\begin{figure}[h]
\centering
\begin{subfigure}[t]{0.5\textwidth}
		\centering
		\includegraphics[width=1.0\textwidth]{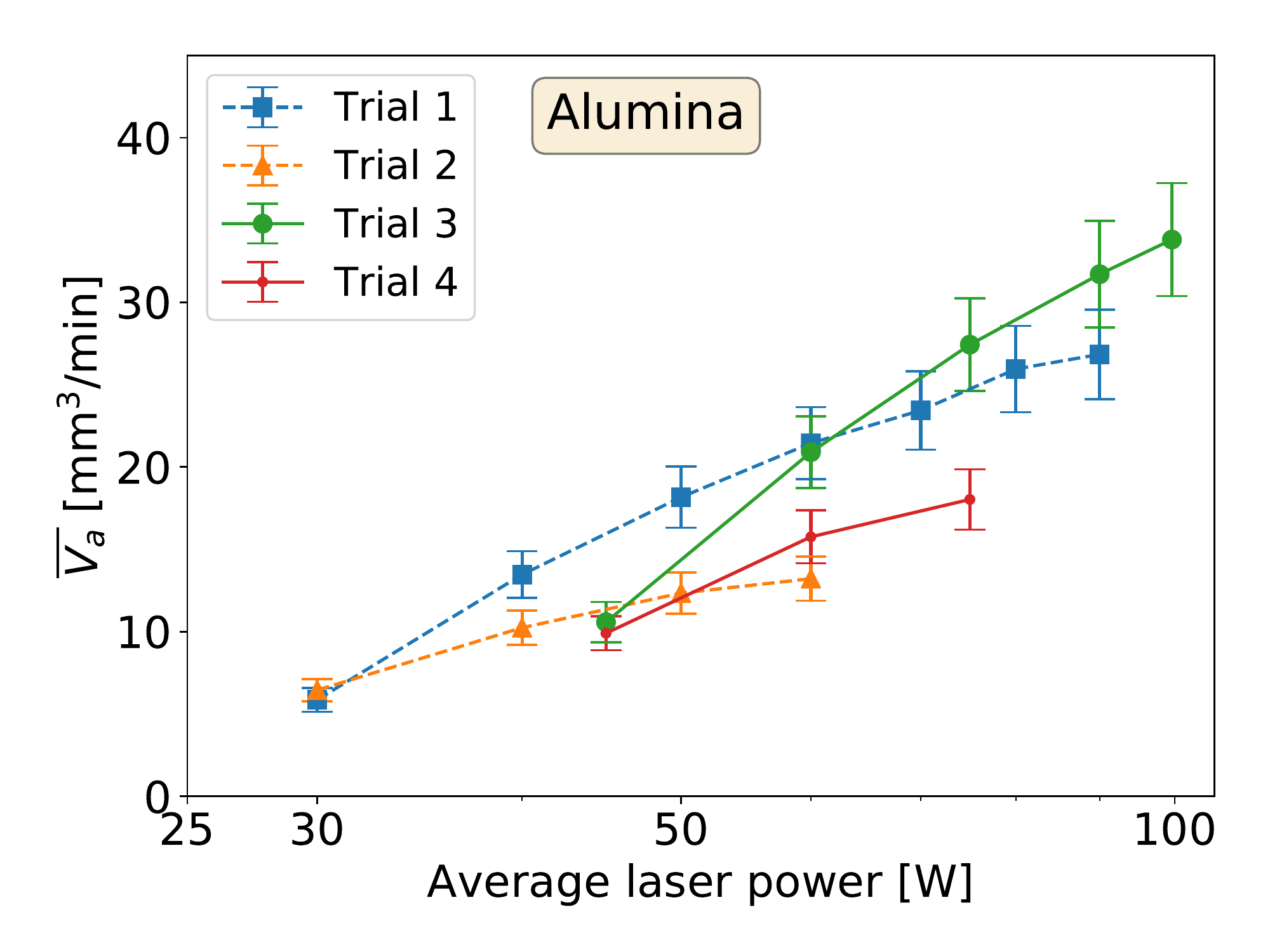}
		\caption{}		
	\end{subfigure}%
\begin{subfigure}[t]{0.5\textwidth}
		\centering
		\includegraphics[width=1.0\textwidth]{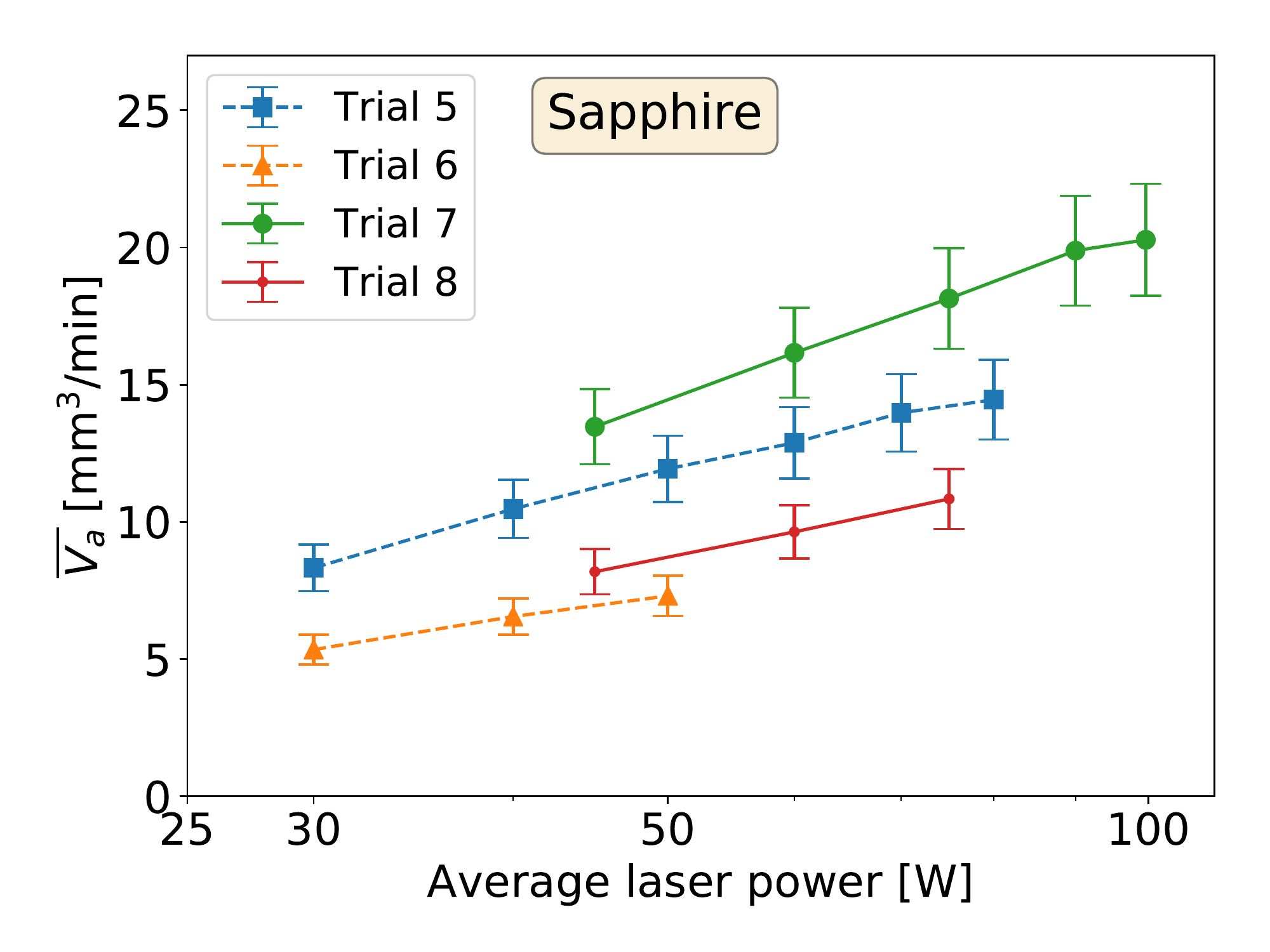}
		\caption{}		
	\end{subfigure}%
\caption{Measured average ablation rate vs. laser power for \textbf{(a)} alumina and \textbf{(b)} sapphire. Error bars include uncertainty in the height measurements and the $\pm 5\%$ uncertainty in estimating the volume removed. Lines between the points are only for visual guidance. 
}
\label{fig:rate_power}
\end{figure}

Figure~\ref{fig:rate_power} gives average ablation rate as a function of laser power. The volume removed $\Delta V$ is assumed to be $50\% \pm5\%$ of the bulk volume of a material layer with thickness $H$ and sample area $A_{sample}$, where $H$ is the measured height of the structures fabricated and $A_{sample} = fL^2$ (see Eq.~\ref{eq:ablated_area}).  The value assumed for $\Delta V$ is based on measurements of several actual SWS. The measurements agree with expectations for V-shaped grooves. The ablation time $T_a$ was calculated based on Equation~\ref{eq:ta} and the known laser scan parameters. The data show that an increase in laser power leads to higher average ablation rate, however the increase is not linear and depends on specific laser and scan parameters. The highest average ablation rate measured was 34 and 20~mm$^3$/min on alumina and sapphire. This rate was measured with SWS height $H \approx 900$~\textmu m and $H \approx 750$~\textmu m, respectively, but is also a function of laser power and other parameters.  These rates are an order of magnitude faster compared to values we reported earlier fabricating SWS with similar dimensions~\cite{matsumura2016_ARC}. The improvement was a result of both higher laser power and better optimized scan parameters. 
Another useful figure of merit is the specific average ablation rate $\overline{V_{s}}$ defined as the average ablation rate per unit laser power, i.e.  $\overline{V_{s}} = \overline{V_{a}} / P$, where $P$ is the average incident laser power. We find values of $\overline{V_{s}}$ reaching 0.37~mm$^3$/min/W on alumina and 0.30~mm$^3$/min/W on sapphire; the highest values were obtained with 75~W and 45~W average laser power for alumina and sapphire, respectively. We compare these results to other published data in Section~\ref{sec:discussion}. 

\section{A Model for the Ablation Process}
\label{sec:model}
\subsection{Model Derivation}
\label{sec:model_derivation}

\begin{figure}[H]
\centering
\includegraphics[scale=0.4]{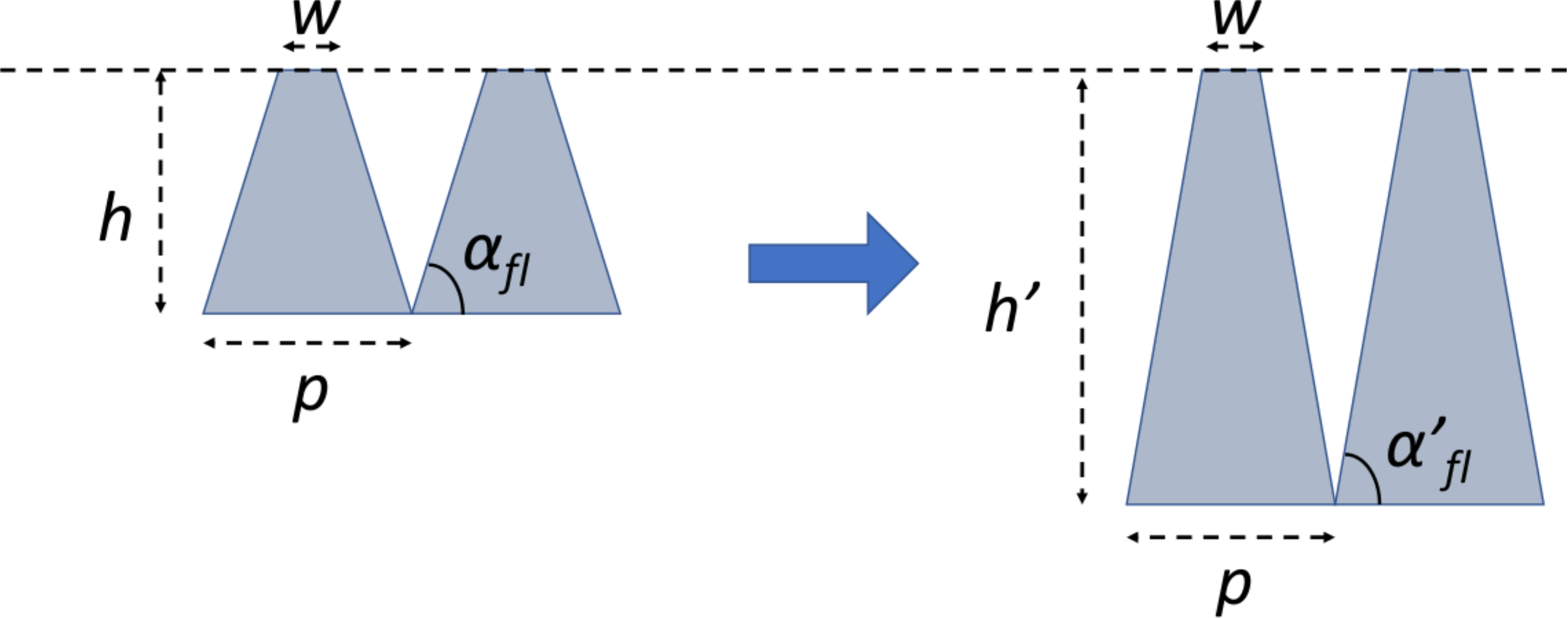}
\caption{In the ablation model we assume that ablation height $h$  increases to $h'$ while the pitch $p$ and the tip width $w$ are constant. The flank angle $\alpha_{fl}$ evolves to $\alpha^{\prime}_{fl}$.
\label{fig:flank} }
\end{figure}

We derive an ablation model that will be used to explain the experimental results. Following the model for ultrashort-pulse ablation first proposed by Furmanski et al.~\cite{Furmanski} and then
developed by others~\cite{Raciukaitis08,Raciukaitis09,Neuenschwander2010,JAEGGI2011164,Neuenschwander2012,NEUENSCHWANDER20141047, Boerner_Hajri_2019}, the instantaneous volume removal rate $v_{rr}$ for a normally incident, Gaussian-profiled beam with a Rayleigh length that is long compared to the ablation depth, with peak fluence $\phi_{0}$, and repetition rate $f$ is
\begin{equation} 
v_{rr} =  \frac{1}{4} f \pi w_{0}^{2} \, \delta \ln^{2} \left( \frac{\phi_{0}}{\phi_{th}} \right)= \frac{\delta \, P }{2 \phi_{0}}  \ln^{2} \left( \frac{\phi_{0}}{\phi_{th}} \right), 
\label{eq:vrr1}
\end{equation} 
where $\delta$ is the absorption length, $\omega_{0}$ is the $1/e^{2}$ beam radius, $\phi_{th}$ is the threshold fluence, and $P$ is the average incident laser power. This model assumes that the laser energy is absorbed according to the Beer–Lambert law, that each pulse is independent of its predecessor, and that the pulses are sufficiently short that heat diffusion during a pulse is much smaller than the absorption length. According to the model there is an optimum peak fluence $e^2 \phi_{th}$ at which $v_{rr}$ per laser power is at maximum~\cite{NEUENSCHWANDER20141047}. Generalizing to an arbitrary incident angle, Boerner et al.~\cite{Boerner_Hajri_2019} have shown that 
the instantaneous volume removal rate is
\begin{equation}
v_{rr} =  \frac{1}{4\cos(\alpha_{i})} f \pi w_{0}^{2} \, \delta \ln^{2} \left( \frac{\phi_{0}\cos(\alpha_{i})}{\phi_{th}} \right)= \frac{\delta \, P }{2 \phi_{0}\cos(\alpha_{fl})}  \ln^{2} \left( \frac{\phi_{0}\cos(\alpha_{fl})}{\phi_{th}} \right),
\label{eq:vrr2}
\end{equation} 
where $\alpha_{i}$ is the incidence angle. 
In reference to Figure~\ref{fig:flank} in which the laser is assumed to illuminate the sample from above, $\alpha_{i} = \alpha_{fl}$. Eq.~\ref{eq:vrr2} can also be intuitively obtained by replacing the peak fluence $\phi_{0}$ in Eq.~\ref{eq:vrr1} with a ``projected peak fluence'' $\phi_{0, proj} = \phi_{0}\cos(\alpha_{fl})$ due to oblique incidence. Refraction is assumed to be normal to the surface regardless of incidence angle. This behavior is expected for metals with high electrical conductivity $\sigma$ for which the angle of refraction $\alpha_r$ satisfies $\tan{\alpha_r} = \sqrt{\frac{2\omega \epsilon_0}{\sigma}}\sin{\alpha_i}$, where $\omega$ is the angular frequency of the incident light~\cite{Orfanidis}. The metallic behavior is justified even for wide-bandgap materials because of the high power, short-pulse ablation~\cite{Boerner_Hajri_2019}.

Using the standard Fresnel expressions we include energy loss due to reflections
\begin{equation} 
v_{rr} = 
  \frac{\delta \, P \left[ 1-R(\alpha_{fl}) \right]}{ 2 \phi_0 \, \cos(\alpha_{fl}) \left[ 1-R(\alpha_{fl}) \right] } \ln^{2} \left\{ \frac{\phi_{0}\, \cos(\alpha_{fl}) \left[ 1-R(\alpha_{fl}) \right] }{\phi_{th}} \right\},
\label{eq:vrr3}
\end{equation} 
where $R$ is the average Fresnel reflectance of the $s$ and $p$ states.

We extend the model to include the entire ablation process of making the SWS. 
We assume that the remaining material after ablation emerges as a 3D symmetrical trapezoid and the ablation proceeds along the triangular geometry shown in Figure~\ref{fig:flank}, with $w$ and $p$ constant. With this geometry, the flank angle and the reflectance $R$ are a function of the varying structure height $h$. Specifically for the flank angle 
\begin{equation} 
\cos(\alpha_{fl})=   \frac{1 }{\sqrt{ 1+4x^{2} } }  \equiv g(h); \,\,\,\,\,\,\, x = \frac{h}{p-w} = \frac{a}{1-w/p}\,\,\, .
\label{eq:angle}
\end{equation} 
In a unit cell with area $p^{2}$ and height $h$ the volume of material removed is
\begin{equation}
V_{removed-cell} = \frac{1}{3}hp^2(2-\frac{w}{p}-\frac{w^2}{p^2}),
\label{eq:removed_volume_cell}
\end{equation} 
therefore over the entire ablation area $A_a$, the volume of material removed is
\begin{equation}
V_{removed} = A_a h k, \,\,\,\, k = \frac{\frac{1}{3}(2-\frac{w}{p}-\frac{w^2}{p^2})}{1-(\frac{w}{p})^2}. 
\label{eq:removed_volume}
\end{equation} 
Values of $w$ and $p$ for the samples give $k$=0.62 and 0.60 for alumina and sapphire, respectively. The volume ablated during a time interval $\Delta t$ corresponding to a layer $\Delta h$ is 
\begin{equation}
k A_{a}\, \Delta h = v_{rr} \, \Delta t = \frac{\delta}{2}  \frac{ P }{ \phi_{0} \, g(h) }   \ln^2 \left\{ \left(\frac{\phi_{0}}{\phi_{th}} \right) [1-R(h)]\,  g(h) \right\}  \Delta t.
\label{eq:volume}
\end{equation} 

We have assumed that the peak fluence $\phi_0$ is constant in the z direction, so that $v_{rr}$ is constant at a given structure height $h$. This assumption is equivalent to assuming an infinitely long Rayleigh length; we discuss this assumption in Section~\ref{sec:model_discussion}.
Collecting the geometrical factors of structures on one side and time variables on the other, assuming that $\Delta h$ is much smaller than $H$, and integrating, we obtain an expression relating the height of fabricated structures to the cumulative fluence 
\begin{equation}
\int^{H}_0 \frac{2k\phi_{0} \,g(h)\, dh }{ \delta\ln^2 \left\{ \left( \frac{\phi_{0}}{\phi_{th}} \right)[1-R(h)]\,  g(h) \right\} } =  \int^{T_{a}}_0 \frac{ P \, dt}{ A_{a} } = \frac{P \, T_{a} }{A_{a} } = F_{cum}.
\label{eq:model1}
\end{equation}
The integrand consists of two known functions $g$ and $R$, two unknown parameters $\delta$ and $\phi_{th}$, and the peak pulse fluence $\phi_{0}$, which is a known parameter depending on pulse energy and spot size. In a given trial, the value of the integral depends on the final height $H$, which is related to the cumulative fluence $F_{cum}$. Both $H$ and $F_{cum}$ are experimentally determined, as shown in Figure~\ref{fig:height_raw_data}. In our analysis we use the experimental data and a least square fit to determine $\delta$ and $\phi_{th}$. 

Alumina and sapphire are polycrystalline and single-crystalline $\alpha$-Al$_2$O$_3$, respectively, with a bandgap of 8.8~eV~\cite{French90}. Theoretical considerations and experimental evidence indicate that in materials in which the bandgap is significantly larger than the incident short-pulse laser radiation the absorption length $\delta$ is a function of the incident intensity~\cite{Kuper:1987ww,doi:10.1063/1.109135,stuart96,Puerto:10,Balling_2013}. Assuming a linear dependence of the effective absorption coefficient $\alpha_{eff}$ on intensity -- thus deviating from linear theory in which $\alpha$  is constant -- we write 
\begin{equation}
\alpha_{eff} = \alpha(1+\gamma I)  \equiv \frac{1}{\delta}.
\label{eq:absorption_two_order}
\end{equation}
Since peak fluence is proportional to intensity, the absorption length $\delta$ is a function of two parameters $\alpha$ and $\hat{\gamma}$ that are to be determined by the data 
\begin{equation}
\delta = \frac{1}{\alpha(1+\hat{\gamma} \phi_{0})}.
\label{eq:delta}
\end{equation}

\subsection{Model Results}
\label{sec:fitting}

\begin{figure}[!h]
\centering
	\begin{subfigure}[t]{0.33\textwidth}
		\centering
		\includegraphics[width=1.0\textwidth]{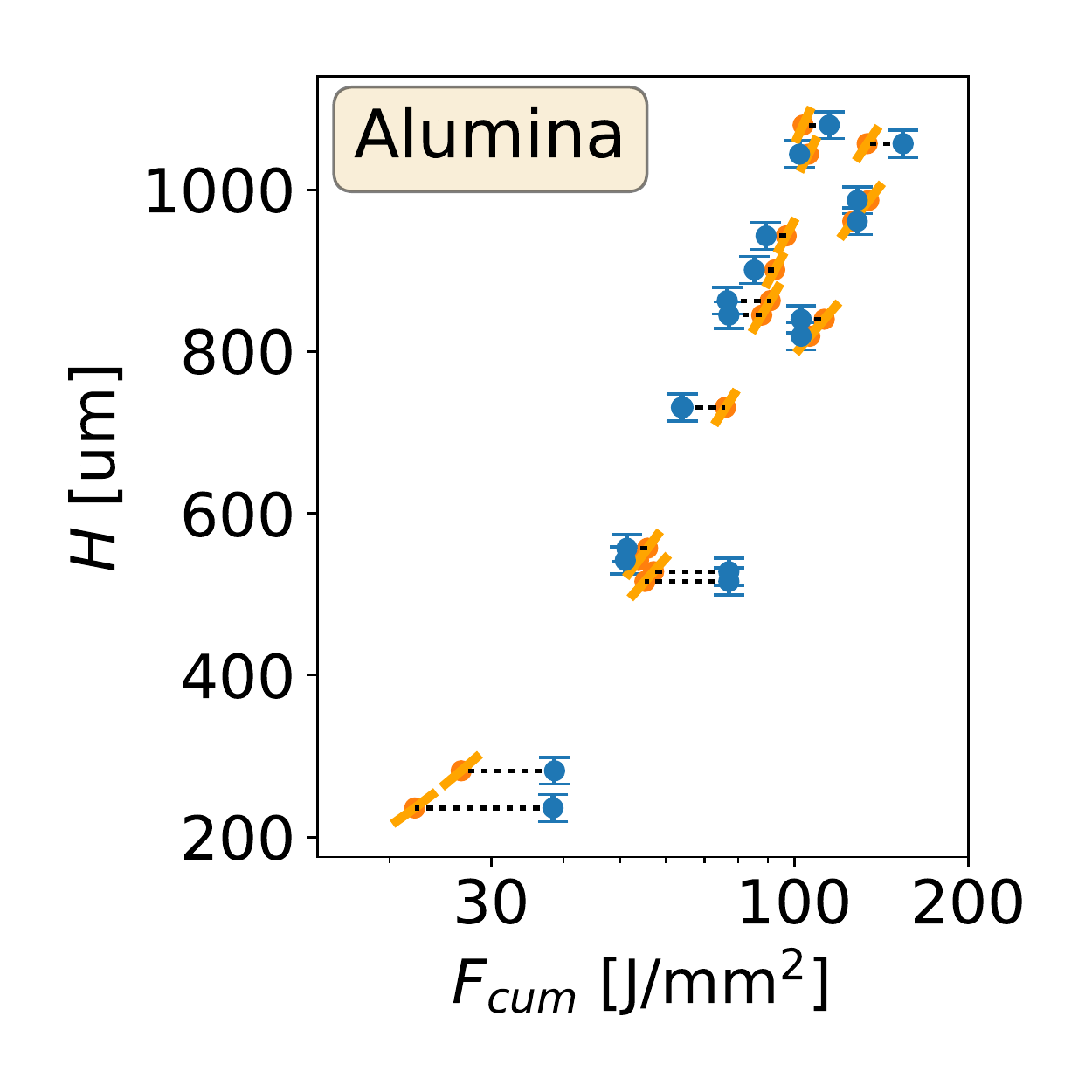}
		\caption{}		
	\end{subfigure}%
		\begin{subfigure}[t]{0.33\textwidth}
		\centering
		\includegraphics[width=1.0\textwidth]{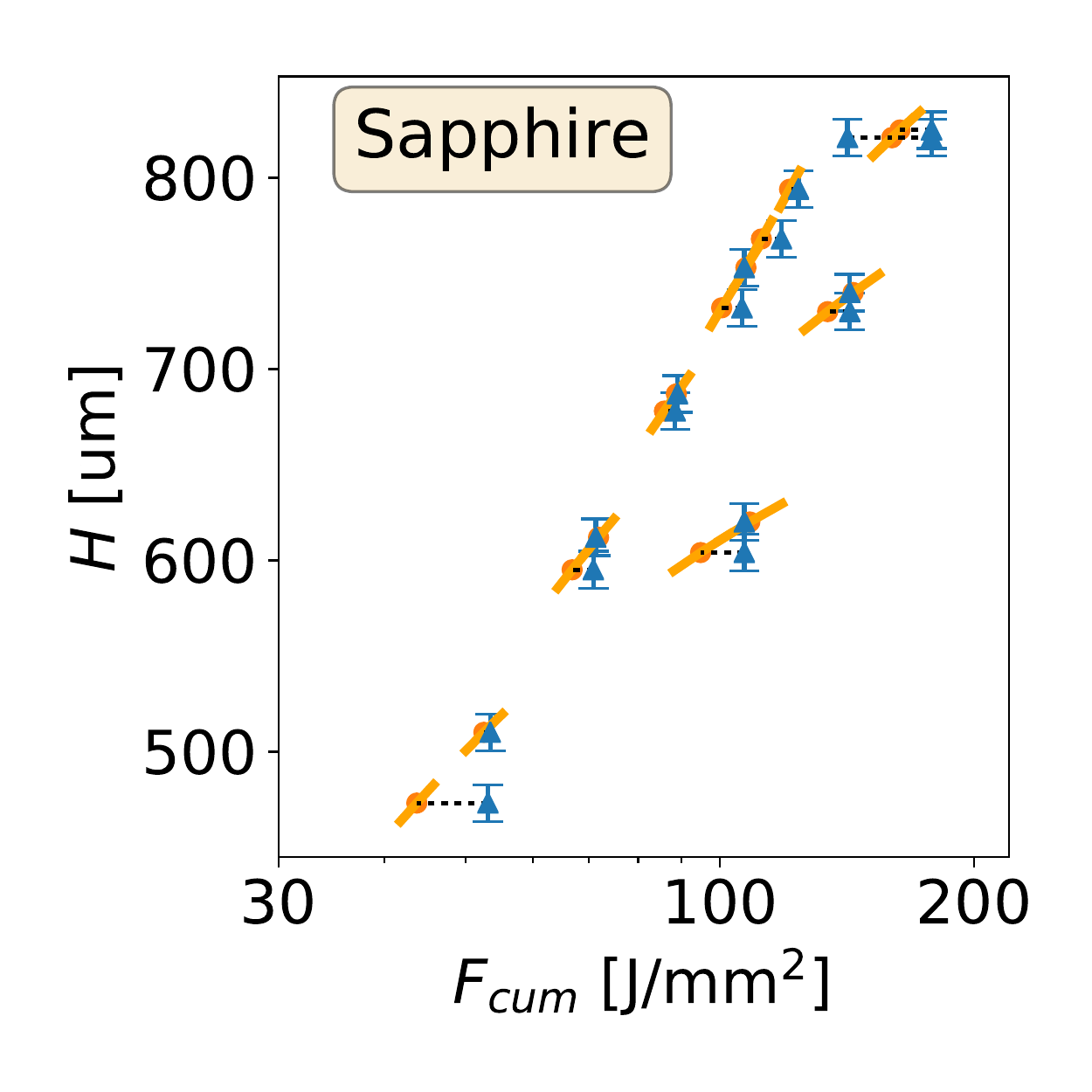}
		\caption{}		
	\end{subfigure}%
		\begin{subfigure}[t]{0.33\textwidth}
		\centering
		\includegraphics[width=1.0\textwidth]{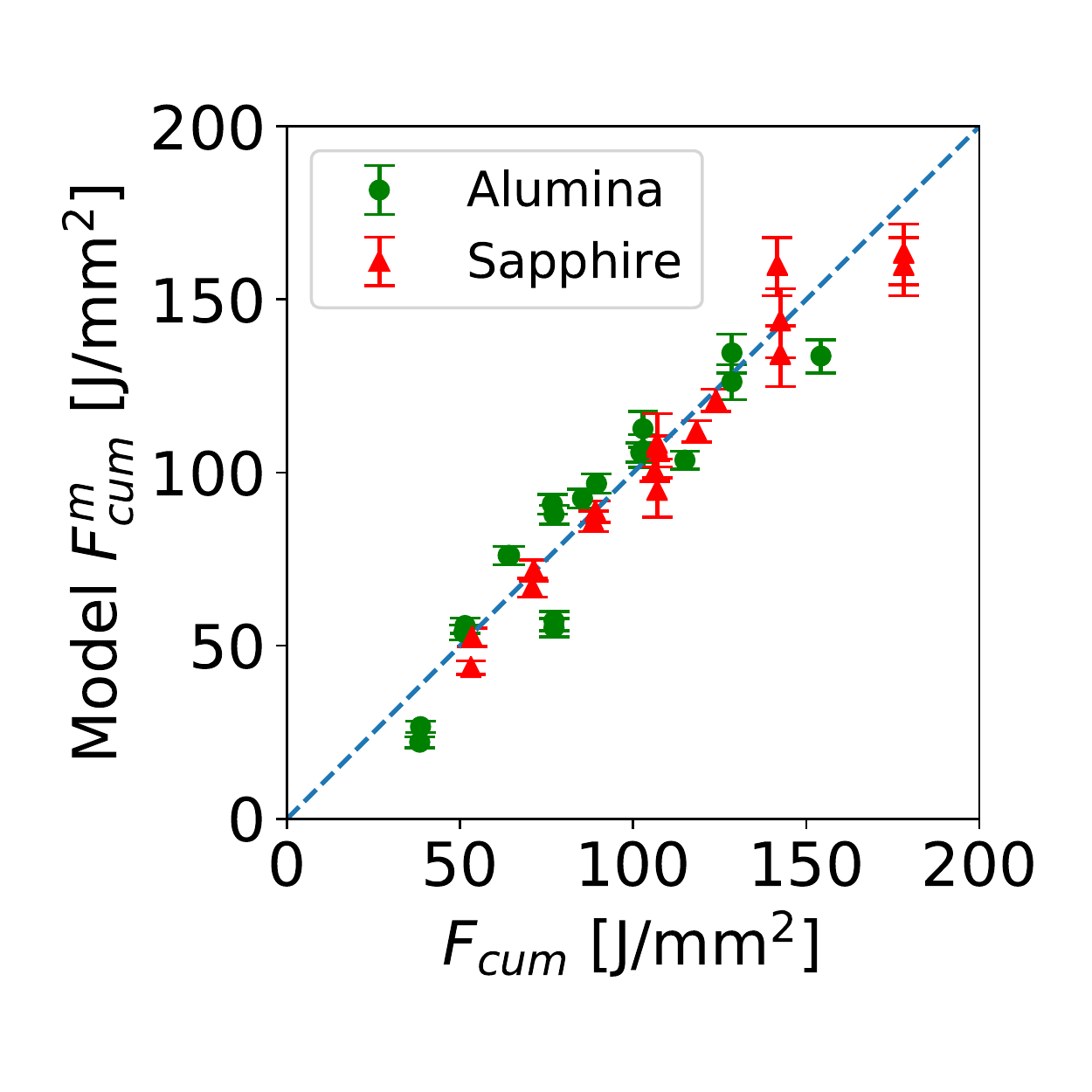}
		\caption{}		
	\end{subfigure}%

\caption{Model data points relating cumulative fluence and height (orange) and comparison to measured data (blue) for \textbf{(a)} alumina and \textbf{(b)} sapphire. Dotted lines connect pairs of experimental and model points that have the same height and peak fluence. Data error bars encode height uncertainties (see Section~\ref{sec:fabricated_structures}). The uncertainty in the calculated cumulative fluence is negligible. Model bars about the model central values indicate $F_{cum}$ values needed to fabricate structures that correspond to the measured $H \pm 1\sigma$ values.  The model gives a one-to-one relation between structure height and cumulative fluence, thus in \textbf{(c)} we plot the model-inferred cumulative fluence as a function of measured cumulative fluence for structures made on both materials. Data errors bars (horizontal axis) are negligible. Model error bars (vertical axis) are projections on the $F_{cum}$ axis from the two left panels. The data fall near the expected slope of 1 (blue dash).   
\label{fig:fits_both} }
\end{figure}

For each material, we used all the measured SWS height and calculated $F_{cum}$ to find the best fit $\phi_{th}$, $\alpha$, and $\gamma$. The values are given in Table~\ref{tab:fit_values} together with uncertainties based on 68\% $\Delta \chi^2$ intervals. With the derived central values for $\phi_{th}$, $\alpha$, and $\gamma$ as inputs, the ablation model of Equation~\ref{eq:model1} was used to {\it predict} cumulative fluence for each of the measured final structure heights. A comparison between the experimentally determined and model-predicted heights as a function of $F_{cum}$ is given in the left and middle panels of Figure~\ref{fig:fits_both}, each for a different material. The RMS differences between the data and the model are 12 and 9~J/mm$^{2}$ for alumina and sapphire, respectively, which represent less than 10\% variance over the 140~J/mm$^{2}$ fluence range of the data. 
An alternative display of the comparison between data and model is given in the right panel where we plot the model-predicted cumulative fluence $F^{m}_{cum}$ for the measured height as a function of the experimental value. 

\begin{table}[!h]
\centering
\normalsize
\def\arraystretch{1.4}
\begin{tabular}{|c|c|c|c|}
\hline
Material & $\phi_{th} $ [J/cm$^2$] & $\alpha$ [$\mu\mbox{m}^{-1}$] & $\hat{\gamma}$ [$\mu\mbox{m}^{-1}/(J/\mbox{cm}{^2}$)] \cr     \hline
Alumina & $2.0\substack{+0.5 \\ -0.5}$ & $2.1\substack{+1.3 \\ -0.9}$ & $-0.005\substack{+0.003 \\ -0.002}$  \cr     \hline
Sapphire & $2.3\substack{+0.1 \\ -0.1}$ & $0.70\substack{+0.48 \\ -0.18}$ & $0.026\substack{+0.012 \\ -0.016}$  \cr     \hline
\end{tabular}
\caption{Model parameters and 68\% confidence intervals obtained from fitting the data.}
\label{tab:fit_values}
\end{table}

\section{Discussion}
\label{sec:discussion}

\subsection{Process Efficiency}
\label{sec:discussion, efficiency}

\begin{figure}[h] 
\centering
\includegraphics[scale=0.5]{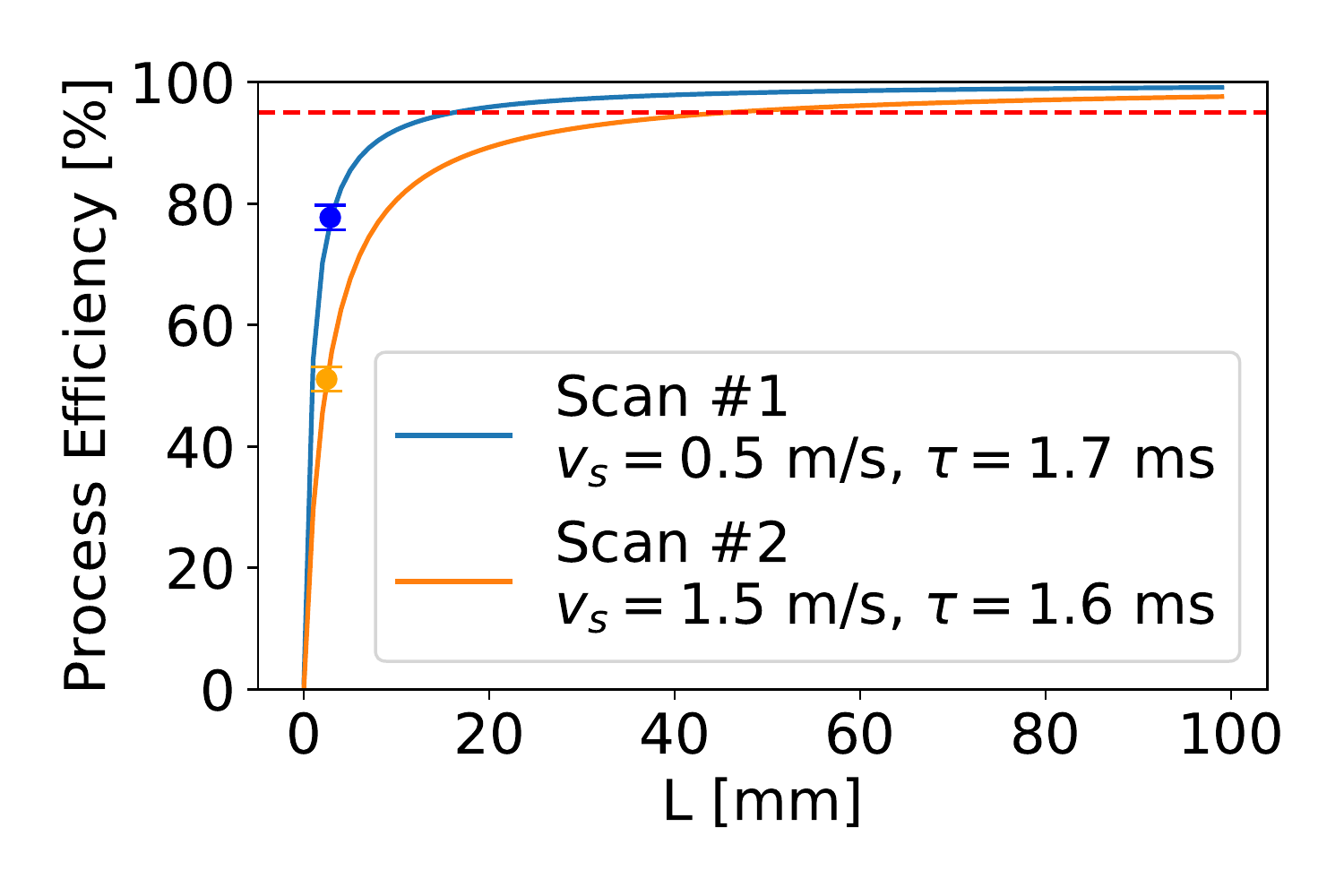}
\caption{Process efficiency as a function of sample size $L$ for two scan settings based on Eq.~\ref{eq:efficiency2} and measured $\tau$. The dots are measured data and the red horizontal line is at $95\%$.}
\label{fig:eefficiency_scan3}
\end{figure}

Equation~\ref{eq:efficiency2} quantifies the dependence of process efficiency on scan parameters. It shows that $\epsilon$ increases as sample size $L$ increases.  Figure~\ref{fig:eefficiency_scan3} shows projections of process efficiencies as a function of $L$ for different scan settings, as well as two of our data points; one with the lowest efficiency and one with the highest. Curves for other data presented in this paper would be located between the two shown curves. 
When fabricating a sample with diameter larger than $\sim$20~mm using laser scan parameters similar to the ones we used for this work, minimizing $\tau$ should be a lower priority, as the process efficiency is near 90\%.  
In all cases, efficiencies are larger than $95\%$ for sample sizes larger than 60~mm. 

\subsection{Volume Removal Rate}
\label{sec:vrr}

The data gave a measured average ablation rate $\overline{V_{a}}$ up to 34 and 20~mm$^3$/min on alumina and sapphire with SWS height $H \approx 900$~\textmu m and $H \approx 750$~\textmu m, respectively. These rates are a  factor  of  34  and  9  higher  than  reported  previously  on  similar  structures~\cite{matsumura2016_ARC}. The highest average rates were both obtained with laser power $P=100$~W. The maximum specific rates were $\overline{V_{s}} = 0.37$ and 0.30~mm$^3$/min/W, and were obtained with $P=75$ and 45~W, for alumina and sapphire, respectively. Schille et al.~\cite{alumina_fast_ablation} reported an ablation rate of 129~mm$^3$/min on alumina using a 187~W ps-laser, giving $\overline{V_{s}}=0.69$~mm$^3$/min/W; Engelhardt et al.~\cite{Engelhardt} reported 205~\textmu m$^3$/pulse with 25~\textmu J/pulse and 200 kHz repetition rate, giving $\overline{V_{s}}=0.49$~mm$^3$/min/W. In those experiments the structures ablated were cavities with flat bottom surfaces and the highest rates were obtained near optimum peak fluence 
(see Section~\ref{sec:model}). Flat surface cavities are conducive to higher ablation rates relative to pyramid-shape SWS because with pyramid shapes (1) the projected fluence is continuously decreasing during ablation, (2) reflection losses are higher due to oblique incidence angles, and (3) it is more difficult to remove ablation debris which scatters some of the incident beam.

\subsection{Ablation Model}
\label{sec:model_discussion}

Fitting the model for the ablation with the data, we find threshold fluence values near 2.0~J/cm$^{2}$ for both alumina and sapphire. Threshold fluence values for these materials reported elsewhere vary between 0.69 and 13~J/cm$^{2}$, and correspond to measurements over a range of wavelengths, pulse durations, repetition rates, and other parameters\footnote{Threshold values were converted to peak fluence if the original results were reported in average fluence. In case of~\cite{ASHKENASI200040} it is not clear whether the fluence reported is peak or average.}~\cite{schutz2016,Furmanski,Boerner_Hajri_2019,ASHKENASI200040,Nieto15,Engelhardt,PERRIE2005213}. Thresholds obtained with laser parameters that are similar to our work~\cite{schutz2016,Boerner_Hajri_2019, Engelhardt} give values between 0.97~J/cm$^{2}$ and 1.4~J/cm$^{2}$, which are within of a factor of two of our results. 

\begin{figure}[h]
\centering
\includegraphics[width=0.6\linewidth]{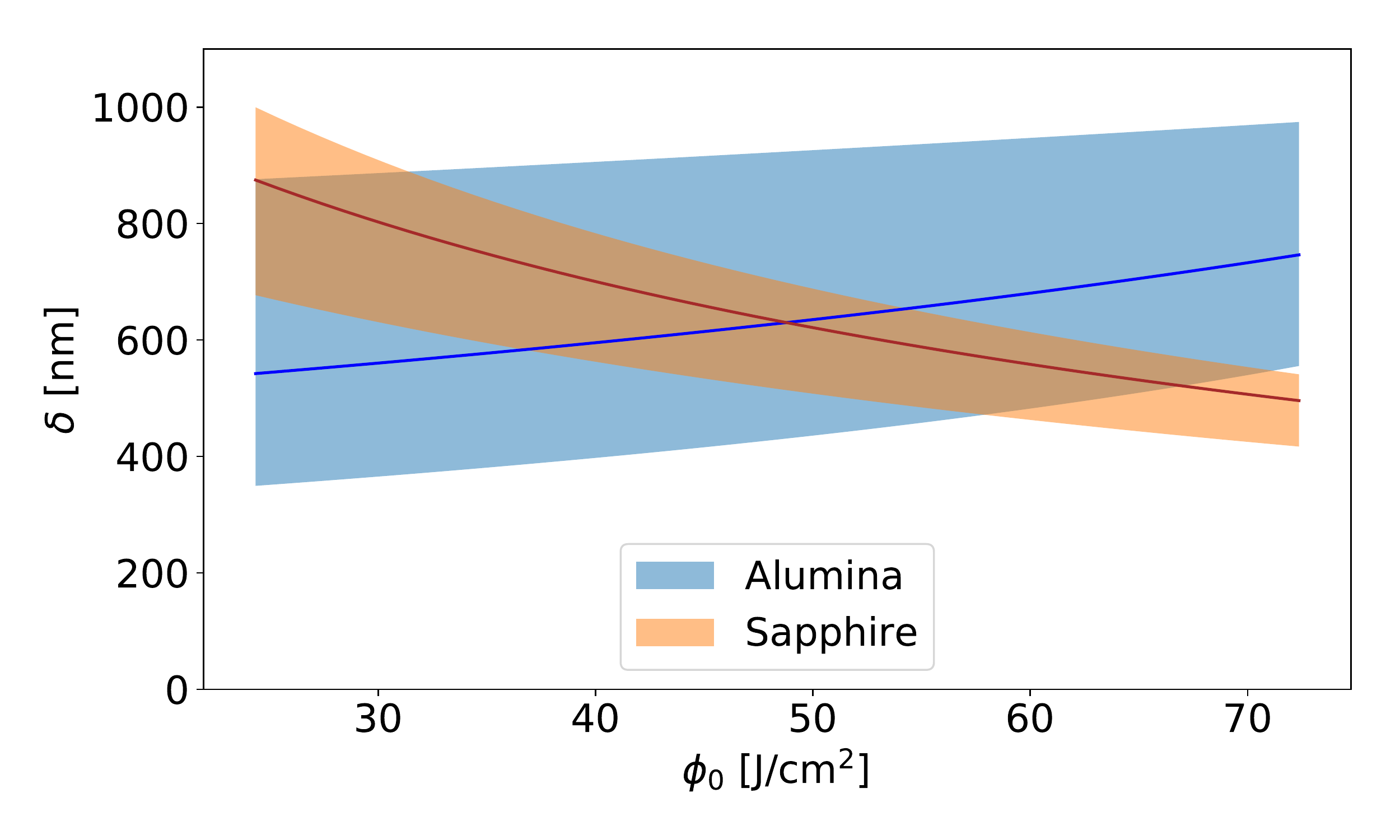} 
\caption{Absorption length as a function of incident peak fluence within the experimental range. The solid curves are calculated based on the best fits given in Table~\ref{tab:fit_values}, while the shaded areas reflect the range of functions allowed given the quoted uncertainties.
\label{fig:fits_penetration}}
\end{figure}

The inferred absorption length $\delta$ as a function of fluence is shown in Figure~\ref{fig:fits_penetration}. The uncertainty intervals encompass $\delta (\phi_{0})$ functions that were determined using pairs of values $\alpha$ and $\gamma$ within their common 68\% uncertainty area; see Table~\ref{tab:fit_values}.
Within uncertainties the data for alumina is consistent with a constant $\delta \simeq 650$~nm, as well as with $\delta$ that mildly increases with fluence.  Furmanski et al.~\cite{Furmanski} reported a constant value of 310~nm on alumina with fluence between 3~J/cm$^2$ and 37~J/cm$^2$. The data for sapphire are consistent with a decrease of $\delta$ with fluence. Boerner et al.~\cite{Boerner_Hajri_2019} reported a constant $\delta=118$~nm for sapphire with fluence between 4~J/cm$^2$ and 18~J/cm$^2$. When we fit our data to a constant absorption length $\delta=1/\alpha$, the RMS difference in cumulative fluence between data and model increases by a factor of 1.1 and 1.3 to 13~J/mm$^{2}$ and 12~J/mm$^{2}$ for alumina and sapphire, respectively; the values for $\delta$ obtained are 770 and 500~\textmu m for alumina and sapphire, respectively.

According to Stuart et al.~\cite{stuart96}, during avalanche ionization the absorption coefficient is proportional to the density of conduction-band electrons excited by the laser beam. Thus a decrease in absorption length with fluence may indicate an increase in production of conduction-band electrons. Alternatively, the observed decrease could be due to other effects not included in the model for the ablation rate such as plasma shielding~\cite{doi:10.1063/1.354325} or debris formation~\cite{VILAR2014313}, both of which can be important especially at high fluence and high repetition rate~\cite{doi:10.1063/1.354325,VILAR2014313,Shaheen_2015}. 

The model presented and quantified through Equation~\ref{eq:model1} simplifies many complex details of the ablation process. Simplifications include ignoring heat diffusion and treating the interaction of each pulse with the material independently from the previous pulse; ignoring the interaction of the incident beam with the ablation plume; and ignoring the redeposition of debris. Even within the framework of the model, it is an approximation. When ablation in the grooves begins, the removed part resembles a trapezoid, not triangles; the possibly complex surface morphological changes are simplified to the progression of a simple geometrical structure; and we assumed an essentially infinite Rayleigh length. This last assumption is justified because the majority of the ablated material was within one Rayleigh length from the focus position, i.e. -0.75 mm $\pm$ 0.54~mm.

Yet, despite its relative simplicity, the model relates total height $H$ to cumulative fluence with RMS of $\sim$10~J/mm$^{2}$ over a range near 150~J/mm$^{2}$ suggesting that it can provide reasonable guidance for future implementations. To compare, when we fit the data assuming vertical-cavity geometry, in which the flank angle $\alpha_{fl}=90^{\circ}$ during the entire ablation process and the prefactor on the left hand side in Eq.~\ref{eq:volume} is 1, the RMS difference between data and model increases by a factor of 1.3 and 2.7 to 15~J/mm$^{2}$ and  24~J/mm$^{2}$, for alumina and sapphire, respectively.

\begin{figure}[h]
\centering
	\begin{subfigure}[t]{0.5\textwidth}
		\centering
		\includegraphics[width=1.0\textwidth]{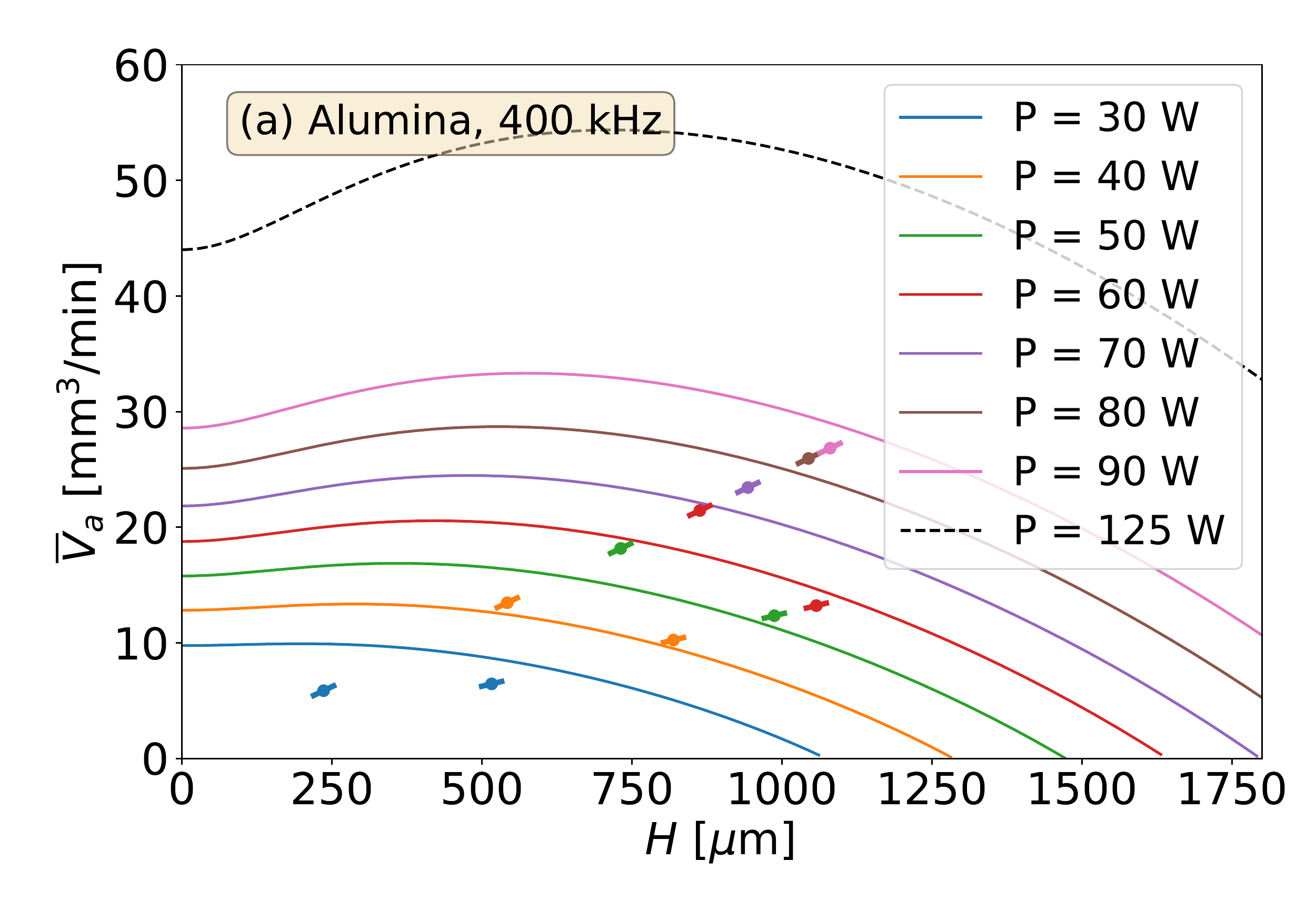}
		\caption{}		
	\end{subfigure}%
	\begin{subfigure}[t]{0.5\textwidth}
		\centering
		\includegraphics[width=1.0\textwidth]{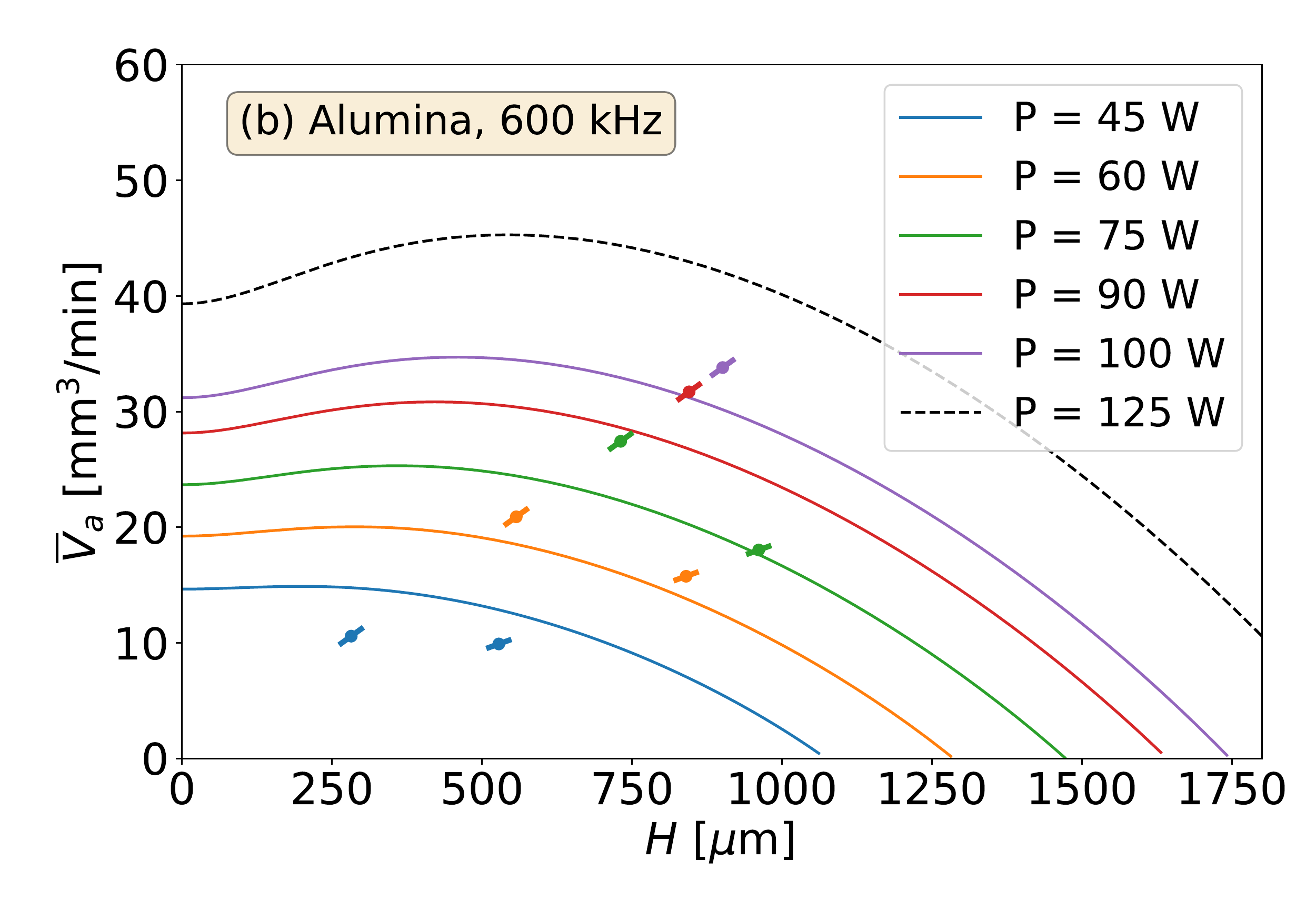}
		\caption{}		
	\end{subfigure}

	\begin{subfigure}[t]{0.5\textwidth}
		\centering
		\includegraphics[width=1.0\textwidth]{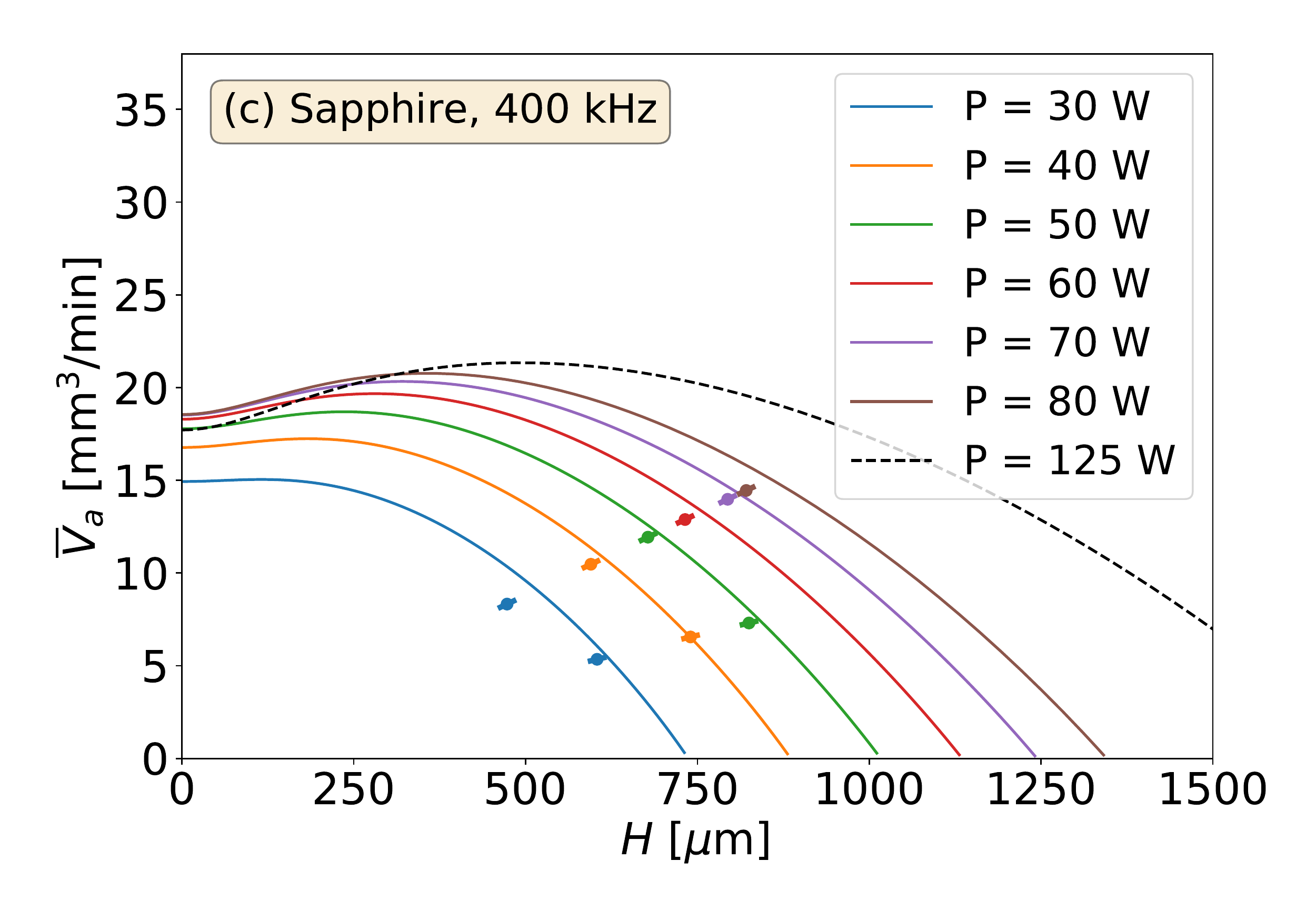}
		\caption{}		
	\end{subfigure}%
	\begin{subfigure}[t]{0.5\textwidth}
		\centering
		\includegraphics[width=1.0\textwidth]{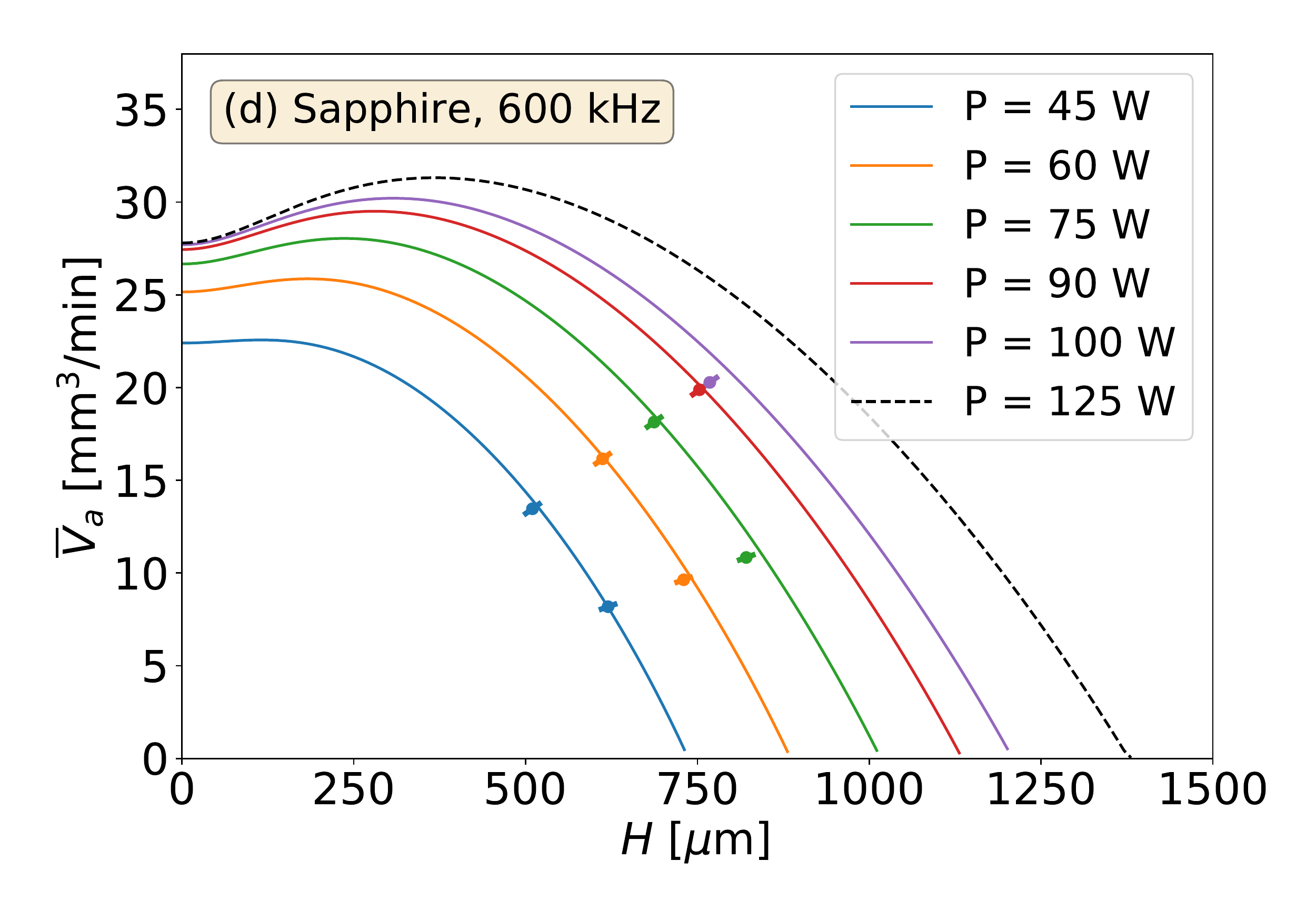}
		\caption{}		
	\end{subfigure}%
\caption{Average ablation rates as a function of the final structure height (lines) predicted based on Eq.~\ref{eq:model1} for different laser powers, and data (points) from trials 1--8. Bars near data points indicate the ranges of $\overline{V_{a}}$ predicted given the uncertainty in $H$. We include a line for $P=125$~W (dash) to indicate the trend with further increase of power. 
\label{fig:rate_height} }
\end{figure}

The model can be used to provide qualitative and quantitative predictions of average ablation rate for any desired structure height in the following way: 
\begin{equation}
\overline{V_{a}}(H) = \frac{\Delta V(H)}{T_a(H)} = \frac{ f_{V} A_{sample} H } {A_a F_{cum}(H)/P } = \frac{1}{(1-w^2/p^2)}\frac{ f_{V} H P } { F_{cum}(H) } , 
\label{eq:rate_model1}
\end{equation}
where for $\Delta V(H)$ we assume that a fraction $f_{V}$ of the volume is removed, and we expressed $T_{a}$ using Equation~\ref{eq:model1}.  The fraction $f_{V}$ and the model predicted cumulative fluence $F_{cum}(H)$ depend on the geometry being ablated and parameters of laser and scan. In Figure~\ref{fig:rate_height} we plot $\overline{V_{a}}(H)$ for different average laser powers $P$ assuming the laser and scan parameters used for trials 1--8,  $f_{V}=1/2$, and $w$=70 and 82 \textmu m for alumina and sapphire, respectively. For a given power  $\overline{V_{a}}(H)$ increases to an optimal rate and then decreases as structure height $H$ increases. This is because at the beginning of the ablation the projected peak fluence $\phi_{0, proj} =\phi_{0}\cos(\alpha_{fl})$ is typically higher than the optimum peak fluence $e^2 \phi_{th}$. The maximal instantaneous removal rate $v_{rr}$ per unit power is only reached after the absorbed projected peak fluence drops to the optimal value as structure height $H$ increases. Material ablation terminates when the absorbed projected peak fluence drops below the threshold fluence. At given power, higher aspect-ratio structures, i.e. with higher $H$ or smaller $p$, have lower $\overline{V_{a}}(H)$ due to smaller projected fluence and larger reflection. Higher laser power increases $\overline{V_{a}}(H)$ for alumina because the absorption length monotonically increases (see Figure~\ref{fig:fits_penetration}), but higher power may decrease $\overline{V_{a}}(H)$ for sapphire, at least for a subset of $H$ values, because absorption length decreases. 

Further improvement in ablation rates for the purpose of making SWS ARC for large optical elements in the millimeter and sub-millimeter wave band require direct measurements of the absorption length $\delta$ and the values of $\phi_{th}$ for the relevant materials, and verification of the ablation model using a range of fluence values, power levels, and geometries.   With constant laser power, further increases in volume removal rate may also be achievable by varying the $z$ position of the beam focus as ablation progresses and by better optimizing the scan. 

 The data give anecdotal evidence for occasional SWS damage at structure heights exceeding $\sim$1~mm and with pulse energies exceeding 200~$\mu$J/pulse. A more systematic study is required to characterize and quantify this effect. 

\section{Conclusions}
\label{sec:conclusion}

We tested a range of ablation parameters for fabricating millimeter-wave SWS structures on alumina and sapphire. We used a 1030~nm picosecond laser that had up to 100~W average power and  achieved average ablation rates of 34 and 20~mm$^{3}$/min with alumina and sapphire SWS heights of 900 and 750~\textmu m, respectively; the aspect ratios of these structures are 2.75, and 2.6, respectively. The highest specific rates achieved were 0.37 and 0.3~mm$^3$/min/W, and they were obtained with laser power of 75 and 45~W, respectively. We demonstrated improvements in average ablation rate of up to a factor of 34 with alumina and 9 with sapphire compared to previously reported rates for making similar structures. with the higher rates, laser-ablating 1~mm tall SWS ARC on a 500~mm diameter optical element should take week, instead of few months.  The significant reduction of processing time makes this technology competitive for broad-band, cryogenically robust, anti-reflection coatings in the MSM astronomy community.

We extended a model for the ablation and compared it with the measured data. We found that despite significant simplifications, the model provides reasonable guidance for the relation between structure height and required cumulative fluence. Over a range of 140~J/mm$^{2}$ in cumulative fluence the RMS differences between the data and the model are 12 and 9~J/mm$^{2}$ for alumina and sapphire, respectively. The best fit values for the absorption length $\delta$, which are in the range of few hundreds of nm, and for threshold fluence $\phi_{th}$ (same table) are comparable with values reported in other publications. 

The model indicates that a primary reason for the higher ablation rates is the increase in laser power. Further optimizations of the ablation process are feasible and thus achieving ablation rates of tens of~mm$^{3}$/min is possible when fabricating structures with heights of $\sim$1~mm.  

\section*{Acknowledgements} 
We acknowledge support provided by Trumpf GmbH \& Co. KG (Ditzingen, Germany); the laser machining trials reported here were done at their laser application center. We thank Christopher Geach for helpful discussions. TM was supported by World Premier International Research Center Initiative (WPI), MEXT, Japan.

\appendix

\section{Process Efficiency}
\label{sec:process_efficiency_derivation}
The ablation time $T_a$ is the ablation time per unit cell of area $p^{2}$ multiplied by the number of such unit cells. The effective number of unit cells is
\begin{equation}
   N_{c} = \Omega L^2/p^2,
\end{equation}
where the form factor $\Omega$ depends on the shape of the sample and is defined in Eq~\ref{eq:ablated_area}. The total ablation distance within a unit cell is $2n_{lines}p$; a factor of two comes from the scans in both $x$ and $y$ directions. Therefore the total ablation time is
\begin{equation}
T_a = N_{L}N_{c}(\frac{2n_{lines}p}{v_s})
 = 2 \Omega N_{L}  (L/p)  n_{lines}  (L/v_{s}), 
 \label{eq:appendix_Ta}
\end{equation}
which is Equation~\ref{eq:ta}.
 
With the scan strategy described in Section~\ref{sec:samplefab} the transitions between lines occur at the edge of the sample.  The total transition distance per layer including both $x$ and $y$ directions is one half the edge perimeter for a square sample, and the entire circumference for a circular sample. The transition time is
\begin{eqnarray}
T_{trans} &=& \begin{cases}
{N_{L}2L/v_{trans},} & \text{square sample}
\\[4pt]
{N_{L}\pi L/v_{trans},} & \text{circular sample}
\end{cases}
\\
    &=&   N_{L}\widetilde{\Omega}L/v_{trans},
\end{eqnarray}
where $v_{trans}$ is the transition speed, assumed to be constant, and another form factor $\widetilde{\Omega}$ accounts for the appropriate geometrical factor.  
We neglect the transitions when switching between $x$ and $y$ direction scans and between layers. 

To calculate $T_{delay}$ we define a delay time per line $\tau_{delay}$, which includes motion delays of the scanner, delays in computer-scanner communications, programmed shutter delays, and potentially other delays. Then
\begin{equation}
T_{delay}   =  2 N_{L}  (L/p)  n_{lines} \tau_{delay}.
\label{eq:td}
\end{equation}

With these relations the process efficiency is 
\begin{equation}
\epsilon   =  \frac{T_{a}}{T_{p}} = \frac{T_{a}}{T_a + T_{trans} + T_{delay}} = 
  \frac{L/v_{s}}{L/v_{s}+\tau} ,
\label{eq:efficiency3}
\end{equation}
where $\tau$ is an average `parasitic' (= non-ablation) time per line that includes line transitions and other delays, but does not depend on the sample size $L$. The expressions for $\tau$ are
\begin{eqnarray}
\tau(p, n_{lines}, v_{trans}, \tau_{delay})  = \begin{cases}
{\frac{p /v_{trans}}{n_{lines}}+\tau_{delay}} & \text{, square sample}
       
\\[4pt]
{\frac{2p/v_{trans}}{n_{lines}} +\frac{4}{\pi} \tau_{delay}} & \text{, circular sample}.
       \\
\end{cases}
\label{eq:tau}
\end{eqnarray}

\bibliographystyle{elsarticle-num}
\bibliography{references}
\end{document}